\documentclass[11pt,a4paper]{article}
\usepackage{jheppub}
\newcommand{\be}{\begin{equation}}
\newcommand{\en}{\end{equation}}
\newcommand{\bea}{\begin{eqnarray}}
\newcommand{\ena}{\end{eqnarray}}
\newcommand{\lbl}[1]{\label{eq:#1}}

\newcommand{\rf}[1]{(\ref{eq:#1})}

\newcommand{\braque}[1]{{\langle #1 \rangle}}
\newcommand{\gapprox}{%
\mathrel{%
\setbox0=\hbox{$>$}\raise0.6ex\copy0\kern-\wd0\lower0.65ex\hbox{$\sim$}}}
\newcommand{\lapprox}{%
\mathrel{%
\setbox0=\hbox{$<$}\raise0.6ex\copy0\kern-\wd0\lower0.65ex\hbox{$\sim$}}}
\newcommand{\inleft}{%
\mathrel{%
\setbox0=\hbox{$<$}\copy0\kern-0.5\wd0\lower1.1\ht0\hbox{$\scriptstyle{in}$}}}
\newcommand{\inright}{%
\mathrel{%
\setbox0=\hbox{$>$}\copy0\kern-0.5\wd0\lower1.1\ht0\hbox{$\scriptstyle{in}$}}}
\newcommand{\outleft}{%
\mathrel{%
\setbox0=\hbox{$<$}\copy0\kern-0.5\wd0\lower1.1\ht0\hbox{$\scriptstyle{out}$}}}
\newcommand{\outright}{%
\mathrel{%
\setbox0=\hbox{$>$}\copy0\kern-0.5\wd0\lower1.1\ht0\hbox{$\scriptstyle{out}$}}}
\newcommand{\mkd} {m^2_K}

\newcommand{\mpid} {m^2_\pi}

\newcommand{\mtaud} {m^2_{\tau}}

\def\tr{ {\rm tr}\,}
\def\Lag{ {\cal L}}

\makeatletter
\def\captionof#1#2{{\def\@captype{#1}#2}}
\makeatother
\title{
{\boldmath  
First determination of $f_+(0) |V_{us}|$ from a combined analysis of 
$\tau\to K\pi \nu_\tau$ decay 
and $\pi K$ scattering
with constraints from  $K_{\ell3}$ decays 
}}

\author[a]{ V\'eronique Bernard}
\affiliation[a] {Groupe de Physique Th\'eorique, 
Institut de Physique Nucl\'eaire, CNRS/IN2P3,\\ 
Universit\'e Paris-Sud 11,
91406 Orsay, France}
 
\emailAdd{bernard@ipno.in2p3.fr}

\abstract{
We perform a combined analysis of
 $\tau\to K\pi \nu_\tau$ decay  and $\pi K$
scattering with constraints from $K_{\ell3}$ data
using a $N/D$ approach that fulfills requirements from unitarity and
analyticity. We obtain a good fit of the $I=1/2$ $\pi K$
amplitude in the $P$ wave using the LASS data
above the elastic region while in this region data are generated via Monte Carlo using
the  FOCUS results based on $D_{\ell 4}$ decay.  The spectrum and branching
ratio of 
$\tau\to K\pi \nu_\tau$ constrained by  $K_{\ell3}$ decays  are also
well reproduced leading to $f_+(0) |V_{us}|= 0.2163 \pm 0.0014 $. Furthermore, we
obtain the slope of the vector form factor $\lambda_+=(25.56 \pm 0.40) \times 10^{-3}$ while the
 value
of the scalar form factor at the Callan-Treiman point is $\ln C=0.2062 \pm 0.0089$. 
Given the experimental precision  our results are compatible with the  Standard Model. 
}

\begin{document}
\maketitle
\flushbottom

\section{Introduction}

One important issue in the test of the Standard Model (SM) as well as
in various new physics scenarios is the possible violation of the
unitarity of the Cabibbo Kobayashi Maskawa matrix (CKM) as well as the
determination of bounds on it.  With the present very precise
knowledge of the element $V_{ud}=0.97425 \pm 0.00022$ from the
superallowed $0^+ \to 0^+$ nuclear $\beta$ decays \cite{PDG,
  Hardy:2008} a determination of $V_{us}$ allows for such a test
between the elements of the first row
$|V_{ud}|^2+|V_{us}|^2+|V_{ub}|^2=1$.\footnote{Indeed one can safely 
neglect the third element which is very small $|V_{ub}| = (4.15
  \pm 0.49)~10^{-3} $ \cite{PDG}.} There exists several ways of
  extracting the matrix element $V_{us}$. One of them is the
study of leptonic and semileptonic kaon decays.  In the latter the combination
$f_+(0) |V_{us}|$ enters, with $f_+(0) $  the strangeness changing
vector form factor at zero momentum transfer. With the progress on the
theoretical side in determining the radiative corrections  and
isospin breaking effects as well as the progress from the lattice
community a very precise determination of $|V_{us}|$ from these decays
becomes possible. At present a global analysis including results
published by the BNL-E865, KLOE, KTeV, ISTRA+ and NA48 experiments
leads to \cite{Flavia2} $f_+(0) |V_{us}|= 0.2163 (5)$.

Furthermore, it has
been advocated in Refs.~\cite{Bernard:2006gy, Bernard:2009zm} 
that the study of $K_{\ell 3}$ decays in particular
offers another possibility to test the SM through the 
determination of the scalar form factor at the  Callan-Treiman (CT) point. 
Indeed $SU(N_f) \times SU(N_f)$ low-energy theorems  dictate  the value of 
this form factor at that point for $N_f=2$ as well as its soft kaon analog 
for $N_f=3$. Thus a deviation from the value at the CT point, once 
the corrections $\Delta_{CT}$ to the theorems are very precisely known
%\footnote{
%These have been 
%calculated in Chiral Perturbation theory in~\cite{GL, Bijnens:2007, Kastner:2008}. $\Delta_{CT}$ is of the order $10^{-3}$ for the neutral decay mode 
%while due to 
%an enhancement from $\pi^0 \eta$ mixing it is of the order of a few $10^{-2}$  
%for the charged channel.}, 
would be a sign of physics
beyond the SM such as right-handed quark couplings to the  $W$ boson or charged Higgs
effects (see for example the discussion in \cite{Flavia1}). This has triggered
a renewal of activity on the experimental side. Three collaborations,
NA48~\cite{NA48}, KLOE~\cite{KLOE} and KTeV~\cite{KTeV} reanalyzed their
data on $K^0_{\mu 3}$ decays so as to extract the value of the scalar form factor at the CT point.
With the current experimental precision,  NA48 has a 4.5$\sigma$ deviation
from the SM while KLOE/KTeV show a good/marginal agreement with the SM.
However, the NA48/2 experiment has recently released preliminary results
for the form factors for both $K^{\pm}_{e3}$ and $K^{\pm}_{\mu 3}$ decays
which  are now consistent with the results from the two other
collaborations ~\cite{NA48ch}. 
In fact, there seems to be some inconsistencies
in the older measurement from NA48 for $K_L$ \cite{Moulson:2013wi}.

Additional information on the  quantity $f_+(0) |V_{us}|$ as well as on the 
scalar form factor can be gained from the dominant Cabibbo-suppressed $\tau$
decay  $\tau\to K\pi \nu_\tau$. It has been
measured by BaBar~\cite{BaBar} and Belle~\cite{Belle} and studied by several 
groups \cite{jpp}--\cite{Kimura:2012nx}.
%-jpp,mouss,Jamin:2008qg, 
It was shown that adding constraints from $K_{\ell 3}$  yielded a more 
precise result for the low-energy part of the  vector form 
factor \cite{bjp1}. However, at present this decay has  never been 
used to determine $f_+(0) |V_{us}|$, rather this quantity was taken as input and a  determination
of  the mass and  width of the resonances present in the spectrum was performed.
As noticed recently \cite{BBP} one can extract this quantity from
$\tau$ decays and this is
one of the goals of this work using the experimental constraints from
the publicly available Belle spectrum of  $\tau\to K_S \pi^- \nu_\tau$ decay.

%{\bb [I think this is confusing. I would suppress this part.] \it It was 
%noticed recently~\cite{BBP} that, {\bc as in the studies of $K_{\ell3}$ decays}, 
%since the normalization was canceling in the  expression
%of the number of events, $ f_+(0) |V_{us}|$ was not entering the fit to the 
%data and thus could be obtained from the decay width measurement. }

Information on the mass and  width of the 
resonances  contributing to the form factors can also be obtained from other experiments,  e.g.  productions  ones  as well as
the semileptonic decay $D_{\ell 4}$ \cite{Link:2005ge, BaBarDl4}.
Furthermore Watson's theorem relates the 
phase of the scalar and vector form factors to the phases of $K \pi$ scattering
in the elastic region. 
Another aim of this paper is thus to gather all the information one has
from these decays as well as from production experiments following \cite{mouss}
\footnote {In this reference 
LASS data on $K \pi$ scattering were fitted in order to determine all the parameters related to the resonances 
appearing in the determination of $\tau$ decays. However in that work, the quantity  $f_+(0) |V_{us}|$ was an input parameter.}
 in order to 
 get a very precise determination of the normalized strangeness changing vector
and scalar form factors in a way as model independent as possible.
This is mandatory for  a very precise knowledge of $|V_{us}|$
as well as in  searches for beyond-SM CP violation
in $\tau \to K \pi \nu_\tau$. These are pursued by CLEO \cite{Bonvicini:2001xz} and more recently  Belle and BaBar \cite{BelleCP, BABARCP}.

Here we concentrate on the  region  $\sqrt s <  \sqrt s_{\rm cut} \sim 1.65$ GeV. 
Indeed from threshold to 
$s_{\rm cut}$ inelasticities in the $P$ 
waves are mostly saturated by the $K^*(892)$ and the $K^{*}(1410)$ and to some extend by the $K^{*}(1680)$. This allows us to model the vector form 
factor in a rather simple way 
in that region using a coupled channel $N/D$ method. This method fulfilling
the requirement of analyticity and unitarity the result can be matched to a
three-times subtracted dispersion
relation assuming that the vector form factor has no zeros. This will allow us 
%on one hand to check our results and on the
%other hand 
to make direct contact with other works since
these relations are extensively used in the literature,
for comparable
issues as discussed here let us mention the descriptions of
the $\pi\pi$ form factor in  \cite{Pich:2001pj} and of the $K\pi$ ones in~\cite{bjp,bjp1}. 
In the latter the inelasticities coming from the $K^* \pi$ channel as well as
 the sum rules obeyed by the slope and the
curvature which  will be discussed here, were not taken into account.    
% has been done 
 %at infinity 
%down to the physical\cite{bjp} in two ways
%region of  $K_{\el
%l3}$ decays allowing for a joined analysis
% this form factor to  the formalism developed in  Ref.~\cite{Pich:2001pj} for the $\pi\pi$ form factor and applied to the $K\pi$ case in~\cite{bjp}, namely a three-time subtracted dispersion
%relation for the vector form factor. 
%Three subtractions allow to minimize the uncertainty from the region at
%higher energies.  
%The model for the phase is improved compared to the one used
%in \cite{bjp} in two ways.
%First, as we will see, we ensure that the form  factors fulfill sum rules that relate their
%slope and  curvature.  Second, we implement in our model
%the inelasticities coming from the $K^* \pi$ channel. 
The scalar form factor, in its turn, is  described using a twice-subtracted 
dispersion relation 
following \cite{Bernard:2006gy,Bernard:2009zm}. Since  the energy region 
considered here is larger than  the one considered in these works, one 
subtraction more could help taming the effect of the unknown high energy 
region, in the spirit of what is done for the vector form factor. 
However it is not so helpful since a sum rule is constraining the additional parameter in the expression of the form factor. Thus contrary to our 
previous work \cite{BBP} we will perform the fit to the $\tau$
data with the twice-subtracted dispersion relation and study the dependence of our results on the parametrization of the
high energy region. We will also for comparison study the extensively used
scalar form factor determined from a coupled channel method \cite{jop,ElBennich:2009da}. 
For a recent related work on scalar form factors in semileptonic B-decays,
see Ref.~\cite{Doring:2013wka}.

In section 2, we  define the quantities needed in our analysis. Then 
in section 3 we  detail our model. First we  
briefly summarize the $N/D$ method in the one channel case and then generalize  it
to the two coupled channel one for $\pi K$ scattering. We then describe the
vector and  scalar form factors and  establish the sum rules 
they should fulfill.  We  discuss the parameters
of the fit especially  their expected order of magnitude.   
Results
of several joined fits to
$\tau\to K\pi \nu_\tau$ and $\pi K$ scattering data constrained by $K_{\ell 3}$ decays are given in section 4.  We finally  discuss our 
determination of  $f_+(0) |V_{us}|$ and the role played by constraining
its value in the combined fit as well as the value of the curvature of
the vector form factor and conclude.

%%%%%%%%%%%% SECTION 2 %%%%%%%%%%%%%%%%%
\section{\boldmath $\tau\to K \pi \nu_\tau$  and $K_{\ell3}$ decays }

The differential decay distribution of the decay $\tau \to \bar K^0 \pi^- \nu_\tau$
reads
%\bea
\begin{align}
\label{eq:tauspeckpi} 
&{d\Gamma_{K\pi}(s) \over d\sqrt{s}} = { G_F^2  m_\tau^5 \over 48 \pi^3} S_{EW}^\tau  
% (1 + \delta_{EM}^{K \tau} +\tilde 
%\delta_{\rm  SU(2)}^{K \pi})^2   
 |V_{us}|^2 |f_+(0)|^2  I_{K}^\tau(s)~,
 % \hspace{7.cm}
\\
& I_{K}^\tau(s) = \frac{1}{m_\tau^2}
\left( 1- {s\over \mtaud} \right)^2 \left[ 
\left( 1+ {2s\over \mtaud}\right) 
{q^3_{K\pi}(s)\over s}     \vert \bar f_+(s)\vert^2\right.
\left.
+{3 q_{K\pi}(s) (\mkd-\mpid)^2\over 4 s^2} \vert \bar f_0(s)\vert^2
\right]\ ,\nonumber
%\ena
\end{align}
where $s=(p_\pi+p_K)^2$, $G_F$  is the Fermi 
constant, $S_{EW}^\tau=1.0201(3)$ the short
distance electroweak
correction \cite{Erler} and $q_{K \pi}$ the kaon momentum in the rest
frame of the hadronic system, 
\be
q_{K \pi}(s)=\frac{\lambda^{1/2}(s,m_\pi^2,m_K^2)}{2 \sqrt{s}}~, 
%\cdot \theta\left( s- (m_K+m_\pi)^2 \right)~, ,
\label{eq:qkpi}
\en
with the K\"{a}llen's function $\lambda(s,m_\pi^2,m_K^2)= \left(s-(m_K+m_\pi)^2\right) \left(s-(m_K-m_\pi)^2\right)$. $I_{K}^\tau(s)$  probes the energy-dependence of 
the strangeness changing $K \pi$ form factors  
normalized to one at the origin, $\bar f_{+,0}(s) \equiv  f_{+,0}(s)/ f_{+,0}(0)$. The vector form factor is defined as
\begin{equation}
\langle K^0(p_K)\vert \bar{u}\gamma^\mu s\vert \pi^-(p_\pi)\rangle=  f_+(t)\,(p_K+p_\pi)^\mu +
f_-(t)\,(p_K-p_\pi)^\mu \ ,
\label{eq:fkpi}
\end{equation}
with $t=(p_K-p_\pi)^2$, while the scalar form factor $f_0(t)$ is the combination
\begin{equation}
f_0(t)=f_+(t) +\frac{t}{M_K^2-M_\pi^2} f_-(t)~.
\end{equation}

Eq.~(\ref{eq:tauspeckpi})  does not take into account the long distance
electromagnetic and strong isospin-breaking corrections.  These corrections  
introduce  small $s$-dependent factors  multiplying both the terms 
proportional to  the vector 
and the  scalar form factors as well as an additional interference term
between the two form factors not written here. Once the distribution is
integrated they lead to corrections which have been recently evaluated 
\cite{PAC} and are of the order of a few percent. 
Clearly a very precise determination of $|V_{us}|$ requires a very accurate
determination of all the quantities on the RHS of Eq.~(\ref{eq:tauspeckpi})
(as well as a very accurate measurement of  $\Gamma_{K\pi}$),
however at the level of  
accuracy of the data neglecting these corrections is perfectly 
legitimate. 

In order to determine $f_+(0) |V_{us}|$ an observable of interest is
the branching ratio which is obtained by integrating the decay spectrum
\be
  B_{K \pi }=\frac{G_F^2 m_\tau^5 }{96 \pi^3}  S_{ew} \tau_\tau  | f_+(0) V_{us}|^2  I_{K}^\tau \, ,
\label{eq:bkpi}
\en
with $I_K^\tau$ the phase space integral
\be
I_K^\tau = \int_{(m_K+m_\pi)^2}^{m_\tau^2} I_{K}^\tau(s) \frac{d s}{\sqrt s} \, ,
\en
and $\tau_\tau$ the tau life time. 
%{\bc $\tau_\tau= (290.6 \pm 1)$ fs}),
%$1/\tau_\tau=2.26501 \,10^{-12}$ GeV) 
Up to very recently the experimental value from the Belle collaboration was \cite{Belle},
\be
B_{exp} \equiv B[\tau^- \to \nu_\tau K_S \pi^-] 
= ( 0 .404 \pm 0.002~{\rm (stat)} \pm 0.013~{\rm (syst)} )\,\%\,\,\, \, .
\label{eq:bkpiexp}
\en
A  value consistent with this study was reported in
\cite{Ryu:2013lca}, while the new update 
is about  $1 \sigma$  higher   with an improved accuracy \cite{Ryu:2014vpc}
\be
B_{exp} \equiv B[\tau^- \to \nu_\tau K_S \pi^-]
= ( 0.416 \pm 0.001~{\rm (stat)} \pm 0.008~{\rm (syst)} )\,\%\,\,\, \, .
\label{eq:bkpiexpnew}
\en  
For BaBar results see \cite{Aubert:2007jh}.

Similar expressions as given for $\tau \to K \pi \nu_\tau$ hold for
$K_{\ell 3}$ decays, the hadronic matrix elements for these two
processes being related by crossing. In that case $S_{EW}^{K_{\ell 3}}=1.0232(3)$~\cite{Sirlin82}.
Long distance electromagnetic and strong isospin breaking corrections are again small
$ \delta_{EM}^{K_{\ell 3}} = (0.495 \pm 0.110) \%$ for the neutral channel,
$ \delta_{EM}^{K_{\ell 3}} = (0.050 \pm 0.125) \%$ for the charged channel and
$\delta_{\rm  SU(2)}^{K_{\ell 3}} = 0.029(4)$, \cite{Kastner:2008, Cirigliano}.

Here we are interested in the region from threshold to $\sqrt s \sim 1.65$ GeV
which, as already stated, is dominated by two resonances the $K^*(892)$ and 
the $K^{*}(1410)$, the latter decaying predominantly into $K^* \pi$. It  is thus
 legitimate to use a two channel approach to describe $K \pi$ scattering as well
as  $\tau$ decays in that region, the  two most relevant channels being 
$K\pi$ and $K^*\pi$. These channels will be labelled 
\be
1 \longrightarrow K\pi,\quad
2 \longrightarrow K^*\pi~.
\label{eq:label}
\en
This implies that  in a coupled channel description one has not only to consider 
the strangeness changing vector form factor, 
 Eq.~(\ref{eq:fkpi}), but also the vector current matrix element 
\be
\braque{K^{*+}(p_V,\lambda) \vert \bar{u}\gamma_\mu s\vert \pi^0(p_\pi)}=
\epsilon_{\mu\nu\alpha\beta}\, e^{*\nu}(\lambda) p_V^{\alpha} p_{\pi}^{\beta}\,
H_2(t)~ .
\en
Consequenly, as  in \cite{mouss}, the model generates predictions for $\tau$ decaying into 
$K^* \pi$ via the vector current,  the energy distribution of the decay width 
being
\be
{d\Gamma_{K^*\pi}(s)\over d\sqrt{s}} =
{ G_F^2  m_\tau^3 \over 32 \pi^3}   {q^3_{K^*\pi}(s) } |V_{us}|^2  
\left(1 - \frac{ s}{m_\tau^2} \right) \left(1 +\frac{2 s}{m_\tau^2}\right) |H_2(s)|^2 \, .
\label{eq:gamkstar}
\en
One has information, though not very precise, on the integrated rate 
from Aleph, $R^{\rm{Aleph}} (\tau \to K^*(1410) \nu_\tau \to K \pi\pi 
\nu_\tau )  = \left (1.4^{+1.3 \, \,+0.0}_{-0.9 \, \,-0.4 } \right) \times 10^{-3}$ where 
the first uncertainty comes from the fit to the $K \pi$ invariant mass, while 
the
second uncertainty arises from the possibility for the $K^*(1410)$ to 
decay into $K \eta$ \cite{Barate:1999hi}.

%%%%%%%%% SECTION 3 %%%%%%%%

\section{Model}

\subsection{Vector channel}

Unitarity relates the imaginary part of the vector form factor to  the $K \pi$  scattering amplitude
in the $J$=1 channel. We will thus first describe this scattering in a two channel approach using the $N/D$ 
method.

\subsubsection{\boldmath $N/D $ description of $K\pi$ scattering: one channel case}
Let us consider the partial wave amplitude with total angular momentum one, and more specifically the quantity  $T^1(s)$
which has the proper behavior at threshold, i.e. it vanishes as $q_{K \pi}^2(s)$.   $T^1(s)$ has two kind of cuts, the 
right-hand cut required by unitarity 
\be 
\left({\rm Im} T^1(s)\right)^{-1}=-q_{K \pi}^2(s) \rho(s) \, \, , ~~\rho(s)=\frac{q_{K \pi(s)}}{8 \pi \sqrt{s}} \theta (s-s_{\rm th})~,
\en
and the  unphysical ones from crossing symmetry. In our case the latter
comprise a left hand cut and a circular one in the complex $|s|$ plane for $|s|=m_K^2-m_\pi^2$.  A standard way to determine
the T-matrix using the knowledge of these cuts
is the $N/D$ method, where the partial wave is expressed as the ratio
\be 
T^1(s)=\frac{N(s)}{D(s)}~,
\en
with $D(s)$ encoding the  right-hand cut and $N(s)$ the unphysical ones.  
In the  phenomenological application used here, it should be safe to neglect 
the latter as a first approximation. Indeed, it has been shown in 
\cite{Oller:2000ma} that considering them in a perturbative manner should be 
realistic in the physical region. Also, tadpoles and loops in 
crossed channel are 
soft contributions which will be reabsorbed in some low energy constants. Hence in the zeroth order approximation,  $N(s)=1$
and all the zeros of $T^1$ will be poles of $D(s)$. 
The most general structure of the T-matrix with the unphysical  cuts
neglected thus reads, see \cite{Oller:1998zr} for more
details:
\bea
T(s)&=&\frac {1}{D(s)}~, 
\nonumber\\
D(s)&=&-\frac{ (s-s_0)^2}{\pi} \int_{s_{\rm th}}^\infty d s'\frac{ q_{K \pi}^2(s')  \rho(s')}{(s'-s)(s'-s_0-i\epsilon)^2} +c_0 +c_1 s+\displaystyle\sum\limits_i
\frac{ R_i}{s-s_i}~.
\ena
The poles in $D(s)$ referred to as CDD poles \cite{CDD}  either can be linked to  
particles (resonances/bound states)  with the same quantum numbers as those of the 
partial wave amplitude or enter to ensure the presence of zeros of the
amplitude required by the  underlying theory such as Adler zeros.  

Splitting the two constants $c_0$ and $c_1$ into a leading and a subleading part
(we will discuss in more detail how we define leading and subleading 
in the next section)
\be 
c_i =c_i^{{\rm lead}} +  c_i^{{\rm sub}}~,
\en
one can write 
\bea
T^{{\rm lead}} (s)&=& \left(c_0^{{\rm lead}}+c_1^{{\rm lead}} s +\displaystyle\sum\limits_i
\frac{ R_i}{s-s_i}\right)^{-1}
\nonumber\\
 g(s)&=& c_0^{{\rm sub}} +  c_1^{{\rm sub}}  s-\frac{ (s-s_0)^2}{\pi} \int_{s_{\rm th}}^\infty d s'\frac{ q_{K \pi}'^2  \rho(s')}{(s'-s)(s'-s_0-i\epsilon)^2} \, .
\label{eq:g}
\ena
One thus finally gets the basic equation for the T-matrix:
\be
T(s)= \left(1/T^{{\rm lead}} (s) + g(s) \right)^{-1} 
%= \left(1 +G(s)T^{{\rm lead}}(s) \right)^{-1}\,T^{{\rm lead}} (s)
\label{eq:Tmatrix}.
\en 
Writing
$K^{-1}= (T^{{\rm lead}})^{-1}+ {\rm Re} \,  g$, one recovers
the well known K-matrix approach.

%This part I would completely erase unless you really want to keep it. In this section we are quite 
%general. In the next we mostly say what is written here, but may be it
%was not written well enough so I rewrite it a bit.  The Green's function $ g(s)$ unitarizes the leading $T$-matrix. In
%  the description based on the appropriate effective theory of QCD,
%  $ g(s)$ is obtained from a bubble loop-integral in the $s$
%  channel, as we discuss in the next section. Our final function $ g(s)$ will therefore be proportional to
%  this loop integral, which respects the dispersive representation of
%  Eq.~(\ref{eq:g}).

\subsubsection{Resonance contributions to $K\pi$ scattering}

Before generalizing to the two channel case, let us discuss what we mean by 
leading and subleading order. In the region of  interest here, pions and kaons are
not the only relevant degrees of freedom.  Resonances have to be taken into 
account explicitly. It thus seems natural to  use  the framework of Resonance Chiral 
Theory (R$\chi$PT)~\cite{Ecker:1988, Ecker:1989}. 
This scheme developed in the mesonic sector incorporates
Goldstone bosons and resonance fields within a Lagrangian approach. 
It is based on Large $N_c$ arguments and uses short distant constraints and
OPE results (note that we do not discuss
the problems with a consistent power counting for loop graphs in such an approach here). 
At present, 
most applications have been done at  tree level but some issues related to 
the next-to-leading order which involves complicated one loop
calculations in a non-renormalizable theory have 
already been addressed, for references see \cite{Portoles:2010yt}. The Resonance Chiral Theory Lagrangian
is given by a sum of two terms 
\be
\Lag_{{\rm R}\chi{\rm PT}}=\Lag_{\chi{\rm PT}}+\Lag_R~,
\label{eq:Lagchi}
\en
where $\Lag_{\chi{\rm PT}}$ is the Chiral Perturbation Theory ($\chi$PT) Lagrangian up to a given chiral 
order  but with Low Energy Constants (LECs) 
different from the ones when no resonance terms are present,  whereas $\Lag^R$ is the
part of the Lagrangian  describing the  resonances. 
Consequently, our leading order constants $c_i^{\rm lead}$ will contain the 
contributions
from this Lagrangian at tree level $i.e.$ the leading large $N_c$ contributions
while the $1/N_c$ ones (loops plus subleading tree level contributions) will 
be given by the $c_i^{\rm sub}$  terms. 

Let us concentrate on the resonance part.  
There are two resonances below $\sqrt s \sim 1.65$ GeV: the $K^*(892)$  and 
the $K^*(1410)$. However  the $K^*(1680)$ is rather close and is rather broad, thus it 
can affect 
the description of the close-by region. We will thus consider these
three resonances here. 
%We will not only 
%consider the $K^*(892)$  and the $K^*(1410)$ which contributes 
%predominantly but also the $K^*(1680)$. Indeed even though  it is located at
%the upper end of the region we are interested in, it is rather broad and thus 
%can affect 
%the description of the close-by region. In fact   
%it will turn out to be necessary to include it  in order  
%to get a good description of $K \pi$ scattering 
%up to $\sim 1.65$~GeV. 
%The $K^*(1410)$ decays predominantly into $K^* \pi$, 
%its experimental decay branching ratio in this channel being larger than
%40$\%$ at 
%the 95$\%$ confidence level while it is only $(6.6 \pm 1)$$\%$ in the $K \pi$ channel.
%The branching ratios of the $K^*(1680)$ are about 30$\%$ in the three channels
% $K \pi \, (38.7 \pm 2.5) \%$, $K^* \pi \, \left (29.9^{+2.2}_{-4.7}\right)\%$
%and $K \rho\, \left(31.4^{+4.7}_{-2.1}\right) \%$. 
The experimental decay branching ratios  for the  $K^*(1410)$ and the  $K^*(1680)$ \cite{PDG} are
\bea
 K^*(1410) \, : &&  \quad (6.6 \pm 1)\% \, \,  (K \pi) ,\quad > 40 \%  \,  \,
 (95\% \, \, \rm{confidence \, level}) \, \,  (K^* \pi)   \, \nonumber \\ 
 K^*(1680) \, : &&  (38.7 \pm 2.5)\% \,  \,  (K \pi) , \quad \left (29.9^{+2.2}_{-4.7}\right)\%\,  \,  (K^* \pi) , \quad  \left(31.4^{+4.7}_{-2.1}\right) \% \, \,  (K \rho) \, .
\ena
Since for simplicity we do 
not take into account the $K \rho$ channel, the last  branching ratio of the $K^*(1680)$ 
 cannot 
be obtained in our model. Thus our description of this resonance is not
completely accurate but this should not affect our results in a 
significant way.

There are two ways of describing spin-one particles  in R$\chi$PT (for a general 
review on vector meson chiral Lagrangians, see \cite{Meissner:1987ge}).
Following Ref.~\cite{mouss} , 
we will work in  the vector formalism in which the nonet of the light vector 
mesons are encoded in a matrix $V_\mu$. The  chiral Lagrangian
is given by~\cite{Prades}
%two coupling constants $g_V$ and $\sigma_V$,
\be
\Lag_R=\sum_{i=1}^3\Lag^{(i)}_K + \Lag^{(i)}_V+\Lag^{(i)}_{\sigma}~,
\en  
with
\bea\lbl{lagv1}
&& \Lag^{(1)}_K = {-1\over4}\tr( V_{\mu\nu}  V^{\mu\nu} -2 M_V^2 V_\mu V^\mu)~,
\nonumber\\
&& \Lag^{(1)}_V = {-i\over2 \sqrt{2}} g_V(1)\, \tr( V_{\mu\nu} [u_\mu,u_\nu])~,
\nonumber\\
&& \Lag^{(1)}_{\sigma} ={1\over 2} \sigma_V(1)\, \epsilon^{\mu\nu\rho\sigma} 
\tr( V_\mu \{ u_\nu, V_{\rho\sigma}\} )~.
\ena
Here, $V_{\mu\nu}=\nabla_\mu V_\nu -\nabla_\nu V_\mu$ and $u_\mu$ describes
the light pseudoscalars. 
%
%In ref.\cite{prades93}, for example, the following values are quoted 
%for the coupling constants (based on the extended NJL model)
%\be
%g_V \simeq 0.083,\quad \sigma_V\simeq 0.25  %revised1:\sigma_V value corrected
%\en
%which should serve as a guide as to the orders of magnitudes.
Similarly, the Lagrangian for
an excited vector resonance $V_\mu ^{(n)}$ reads ($n\neq 1$)
\bea\lbl{lagvn}
&& \Lag^{(n)}_V = {-i\over2 \sqrt{2}} g_V(n)\, \tr( V_{\mu\nu}^{(n)} [u_\mu,u_\nu])~,
\nonumber\\
&& \Lag^{(n)}_{\sigma} ={1\over 2} \sigma_V(n)\, \epsilon^{\mu\nu\rho\sigma} 
\tr( V_\mu^{(n)} \{ u_\nu, V_{\rho\sigma}\} )~.
\ena
We have written explicitely the terms which do not involve the quark mass 
matrix and, therefore, 
have exact $SU(3)$ flavor symmetry. This is sufficient for our purposes.

%Experimental results on $K^*$
%resonances indicate that flavor symmetry can sometimes be 
%substantially broken, as in  the case of  the $K^*(1410)$. 
%We do not try to write down all possible Lagrangian
%terms which break flavor symmetry  but eventually will implement 
%such effects at the level of the fits. We also do not consider explicitly the
%possibility of many more terms which involve further derivatives acting
%on the vector or on the chiral fields. Again, such terms which give
%rise to polynomial energy dependence, may be implemented phenomenologically
%as ``background'' contributions in the fits.  
%
%
%\subsection{Resonance contributions to $K\pi$ scattering}
%
%Here we are interested in the region from threshold to $\sqrt s \sim 1.5$ GeV
%which is dominated by two resonances the $K^*(892)$ and the $K^{*}(1410)$, the latter decaying predominently into $K^* \pi$. It thus 
%seems legitimate to use a two channel approach to describe $K \pi$ scattering as well
%as the $\tau$ decay in that region, the two relevant channels being
%\be
%1 \longrightarrow K\pi,\quad
%2 \longrightarrow K^*\pi,\quad
%\en
Using  Eqs.~\rf{lagv1},~\rf{lagvn} 
the resonance contribution to the $T$-matrix has been obtained in Ref.~\cite{mouss}. 
It can be written in a compact form displaying the usual resonance  structure:
\be\lbl{compactres} %revised: different notation here 
T_{ij}^{\rm res} =\sum_n\, {g(n,i) g(n,j) \over M_n^2 - s }~,
\en
with
\bea\lbl{compactres1}
&& g(n,1)= {g_V(n)\over\sqrt{16\pi} }\left( {\sqrt{s}\over F_\pi}\right)^2~,
\nonumber\\
&& g(n,2)= {\sigma_V(n)\over\sqrt{16\pi} } {\sqrt{2s}\over F_\pi}
(1+\delta_{n1})~, 
\ena
and the sum runs in our case  over the three resonances considered here.
 
Implementing $2$-channel unitarity using the $N/D$ method  discussed 
previously the leading order  $T^{{\rm lead}}$-matrix 
%$T^{{\rm {R}}\chi{\rm PT}}$ --> T^lead for consistency
and $g$ are  now  $2 \times 2$  matrices (see Eq.\ref{eq:label} for the
labelling of the channels).
The 
former has the following  general form
\begin{equation}
 T^{{\rm lead}}= \left( \begin{array}{cc}
 a_0 +a_1 s + T_{11}^{{\rm res}} &a_4 \sqrt s + T_{12}^{{\rm res}} \\
a_4 \sqrt s + T_{21}^{{\rm res}} & a_2+ a_3  s + T_{22}^{{\rm res}} 
\end{array} \right)~,
\end{equation}
In the $1 \to 1$ channel $a_0$  and $a_1$ come from the tree level
contributions of the $\chi$PT Lagrangian, Eq.~(\ref{eq:Lagchi}). We refrain
from giving their expressions here but we will
comment more on them  in section \ref{Sec:Results}. 
For the other channels, the  $a_i$ are unknown coefficients.

\begin{figure}[h!]
%\begin{minipage}{.46\linewidth}
%\hspace{-1.5cm}
%\includegraphics[width=7cm,angle=-90]{vecnewb.ps}
%\captionof{figure}{\caption{Representation of the vector form factor}}
%\end{minipage} 
\begin{minipage}{.46\linewidth}
%\hspace{-0.8cm}
\includegraphics[width=6cm,angle=-90]{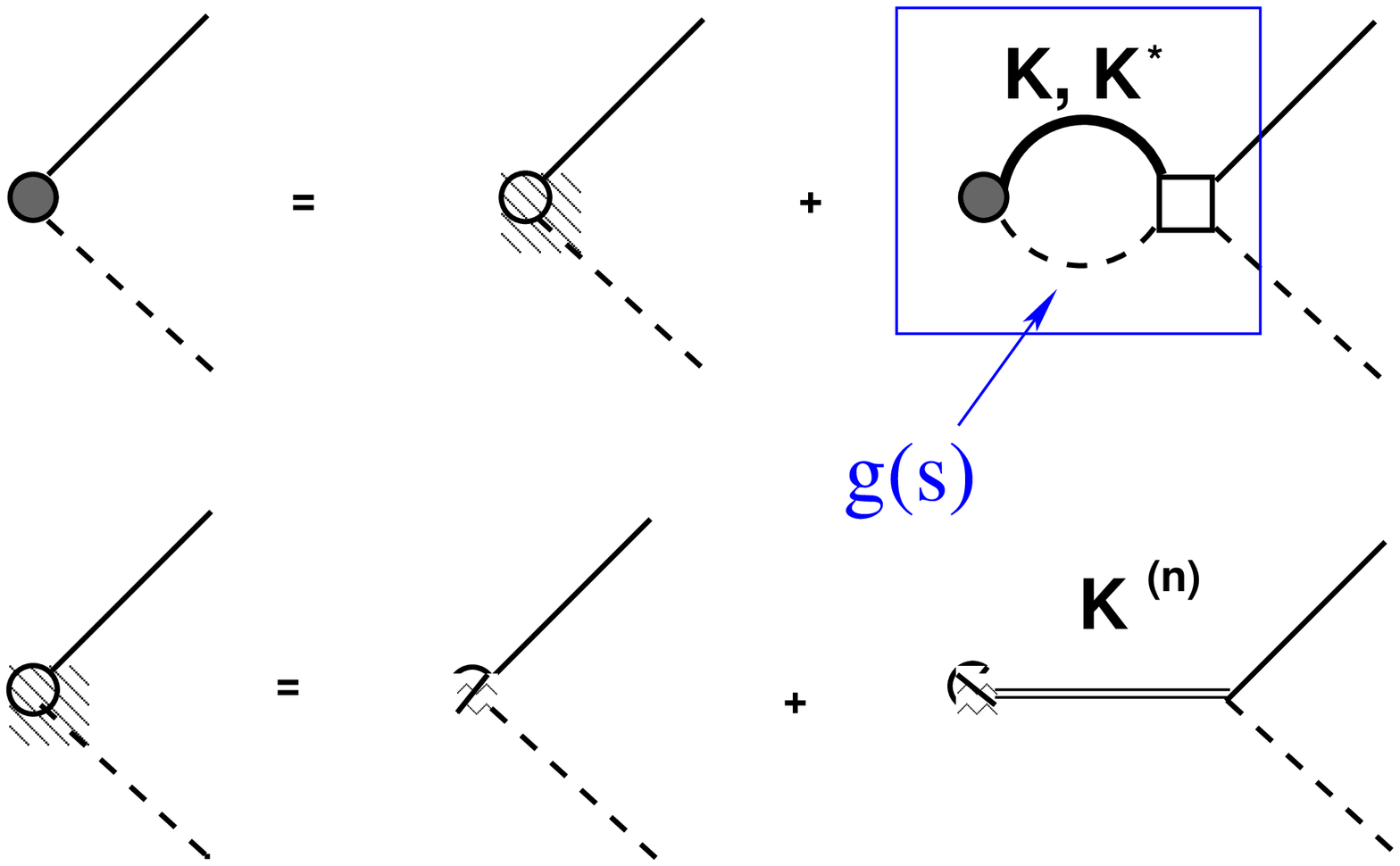}
\end{minipage} 
\begin{minipage}{.46\linewidth}
\includegraphics[width=6cm,angle=-90]{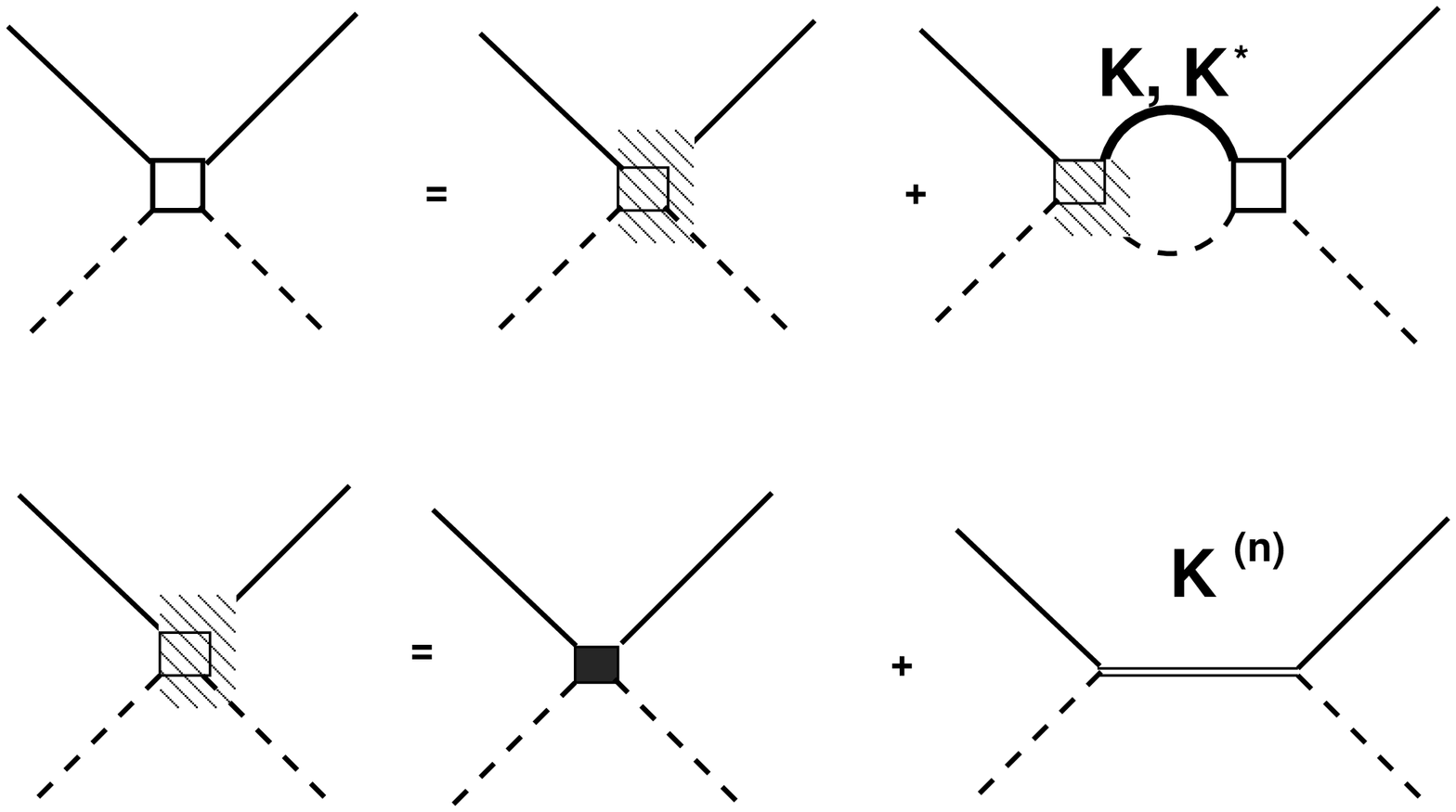}
%\end{minipage}
%\captionof{figure}{\caption{Representation of the $T_{11}$ matrix element.}}
\end{minipage}
\caption{Representation of the $T_{11}$ matrix element (right panel) and of the vector 
form factor (left panel).
The full/dashed external lines represent the kaon/pion respectively. The double line stands for 
one of the three resonances while the thick line in the loops stands for the
kaon or the $K^*$. The fundamental bubble which appears in  the blue
box  on the left panel is described by $g(s)$, Eq.~(\ref{eq:gs}).  }
\label{fig:bubble}
\end{figure}

In our effective theory approach $g$ is the diagonal matrix representing the 
fundamental bubble one-loop-integral 
illustrated in the blue box of Fig.~\ref{fig:bubble}.
It is given by
 \begin{equation}
  g(s) =- \left( \begin{array}{cc}
48 \pi \left(  F_\pi^2 \, H_{K \pi}(s) + l_{K \pi} \right)  & 0 \\
0 & 48 \pi \left( F_\pi^2 \, H_{K^* \pi}(s) +l_{K^* \pi}\right) \end{array} \right)~.
\label{eq:gs}
\end{equation} 
where $H_{a b}(s)$ is the well-known scale-independent function in $\chi {\rm{PT}}$, see Ref.~\cite{GL}
\be
H_{a b}(s)=\frac{1}{F_\pi^2} \left(s M_{a b}^r(s)-L_{a b}(s)\right)+\frac{2}{3 F_\pi^2} L_{ab}^r s~,
\en 
$l_{ab}$ and $L_{ab}^r$ contain the polynomial part of the
loops and the subleading contributions from $\Lag_{\chi {\rm{PT}}}$,
Eq.~(\ref{eq:Lagchi}). Note that  $ L_{ab}^r$ is a scale-dependent quantity
which cancels the scale-dependence from the combination $s M_{a b}^r(s)-L_{a b}(s)$.
As it is written, $g(s)$ respects the dispersive integral, Eq.~(\ref{eq:g}). 
Indeed, it has the same imaginary part and thus can
differ only by  polynomial terms. These can be absorbed into
the parameters $l_{ab}$ and $L^r_{ab}$.
  
The  $K$-matrix approach used in   Ref.~\cite{mouss} can be obtained from these 
expressions defining
$K^{-1}= (T^{{\rm lead}})^{-1}$ and keeping only the imaginary part
of $g$. Similarly to this approach, the $S$-matrix  defined by
\be
S= 1 + 2  \, g \, T~, 
\en
is unitary. 
%and encodes the proper $J=1$ angular momentum barrier factors. 

\subsubsection{Vector form factor}
Following Ref.~\cite{mouss}, we will focus on one of the spatial components of the vector current
and go to the center-of-mass (CM) frame of the meson pair. This allows to project
onto $f_+(s)$ and use unitarity requirements to derive an equation for the vector form factors in
a similar way to what has been done first  
for scalar form factors \cite{moll} and then  for example in \cite{lmei}.
For completeness, we will summarize the argument
here. Defining the matrix of the vector form factors
\begin{equation}
\Gamma (s ) =\left( \begin{array}{c}
 f_+(s) \\
\sqrt s \,  H_2(s) \end{array} \right) ~,
\label{eq:Gamdef}
\end{equation}
unitarity implies the following relation between $\Gamma (s)$
and the $J=1$ $T$ matrix, see Eq.~(\ref{eq:Tmatrix})
\begin{equation}
{\rm Im} \Gamma (s)= T(s) \frac {2 Q^3(s)}{\sqrt s}  \Gamma^*(s) ~,
\end{equation}
with 
%$\tau ={\rm diag} (1, \sqrt t)$ and 
\[Q(s) =\left( \begin{array}{cc}
q_{K \pi}(s) &0 \\
0 & q_{K^* \pi}(s) \end{array} \right) . \]
where $ q_{K^* \pi}(s)$ is defined in a similar way as $q_{K \pi}(s)$, Eq.~(\ref{eq:qkpi}) but with the kaon mass 
replaced by the $K^*$ mass.  
Substituting in the previous equation ${\rm Im} \Gamma(s)$ by $(\Gamma(s)-\Gamma^*(s))/(2 i)$ and
$T(s)$ by its expression, Eq.~(\ref{eq:Tmatrix}), one has
\be
\Gamma(s) =  \left[ I + T^{\rm lead}(s)\,  g(s) \right]^{-1} \left (I+  T^{{\rm lead}}(s) \,  g(s) +T^{{\rm lead}} (s) \frac{ 4 i Q^3(s)}
{\sqrt s } \right) \Gamma^*(s)~.  
\label{G_eq} \\
\en
Taking into account that $T^{{\rm lead}}$ is real and that 
\be
 g^*(s)=  g(s) +  4 i \frac{Q^3(s)} {\sqrt s }~,
\en
one can write
\be
\left( I + T^{\rm lead}(s)\,  g(s) \right) \Gamma(s)= \left( I+ T^{\rm lead}(s)\,g^*(s)\right) \Gamma^*(s)~,
\en
which implies that the quantity $\left( I+ T^{\rm lead}(s)\, g(s) \right) \Gamma(s)$ has no cuts since
the only one which appears in $ g(s)$ and $\Gamma(s)$, the right-hand cut, is removed. 
Therefore one can finally write
\be
\Gamma(s) =  \left[ I+ T^{\rm lead}(s)\, g(s) \right]^{-1}   {\cal{R}}(s)~,  
\label{Gam_eq} 
\en
where  
${\cal{R}}(s)$ is a  matrix of  real functions free of any singularity.  

%J'enleverai toute cette partie en bleue cela me semble redondand.  The unitarized description of the form factors is obtained from $T^{\rm lead}$ and
%  the Green's function matrix $ g(s)$ through the same matrix that 
%  unitarizes the scattering amplitude in Eq.~(\ref{eq:Tmatrix}). The right-hand cut is therefore the same as in the
%  scattering case. Additionally, all the poles of the $T$-matrix
%  are automatically present in the form factors.

We will fix $R(s)$ by requiring  matching to R$\chi$PT obtaining 
\[{\cal{R}}(s)=  \left(\begin{array}{c}
 \left(h_1+ \sqrt {16 \pi}   \sum_n\, g(n,1) F_n s \left( {1 \over M_n^2 - s } +{1 \over s} \right) \right) \\
                                        \\
\sqrt s  \, h_2 + \sqrt {16 \pi} \sum_n\, g(n,2) F_n  {s  \over M_n^2 - s } \\

\end{array} \right)~, \]
with $F_n$ defined by 
\begin{equation}
{\cal L}=- \frac{F_n}{2 \sqrt 2} \langle V_{\mu \nu}^{(n)} f_+^{\mu \nu} \rangle~,
\end{equation}
and
$h_1$ and $h_2 $ are such that the quantities on the right-hand-side of
Eq.~(\ref{Gam_eq}) give the proper normalization of the form factors
at $s=0$. In the expression of the first row of ${\cal{R}}(s)$ we have added
the term $1/ s$ to the resonance contribution.
It has indeed been shown  in
\cite{Ecker:1989} that such a term is required for consistency with QCD when
using a vector field formulation to describe the spin-1 resonances.  

Keeping only one channel  and one resonance and using the relation from R$\chi$PT 
\begin{equation}
F_{K^*} g_V M_{K^*}^2/F_\pi^2  =1
\end{equation} 
with $F_{K^*} \equiv F_1$, Eq.~(\ref{Gam_eq}) reduces to the formula used in Ref.~\cite{bjp}
\begin{equation}
\label{Fpone}
f_+^{K\pi}(s) \,=\, \frac {m_{K^*}^2}{m_{K^*}^2 - s -
\kappa \, H_{K\pi}(s)} \,,
\end{equation}
with $\kappa$ a dimensionful constant.\footnote{ In the last equation we expanded
the coupling of the resonance to $K\pi$, which is proportinal to $s$, around the off-shellness of the  resonance using $s = m_{K^*}^2 +\delta s$~\cite{mouss}.}

\subsection{Dispersive representation of the form factor}

The function $\Gamma(s)$, Eq.~(\ref{eq:Gamdef}) is clearly only a good description
of the form factors up to $\sqrt s \sim 1.65$ GeV, in particular it does not have
the proper behaviour at infinity.
%Either one directly uses Eq.~(\ref{eq:Gamdef}) {\bc for the vector form factor} to 
%determine the $\tau$ decay spectrum or 
This is completely sufficient for our purpose. 
However, since unitarity and the analyticity properties are fulfilled, the vector
form factor can 
be rewritten as a dispersion relation that employs the phase extracted from 
Eq.~(\ref{eq:Gamdef}) supplemented by some parametrization of the phase
at higher energy. In order to 
compare with other works as well as to check our calculation we will match our 
vector form factor to  
a three  times subtracted dispersion relation following~\cite{bjp}, the number 
of subtractions allowing in principle to tame the dependence on the high 
energy part of the phase.  Assuming that the form factor has no zeros one
can write \cite{bjp}:
\begin{equation}
 f_+^{K\pi}(s) \,=\, f_+(0) \exp\Biggr\{ \alpha_1 \frac{s}{M_{\pi^-}^2} +
\frac{1}{2}\alpha_2\frac{s^2}{M_{\pi^-}^4} + \frac{s^3}{\pi}\!
\int\limits^{\infty}_{s_{K\pi}} \!\!ds'\, \frac{\delta_1^{K\pi}(s')}
{(s')^3(s'-s-i\epsilon) }\Biggr\} \,,
\label{eq:ffvdisp}
\end{equation}
where the phase of the form factor $\delta_1^{K\pi}(s)$ in the region
from threshold to $\sqrt s =1.6$~GeV  is  obtained from Eq.~(\ref{Gam_eq}). One has
\begin{equation}
\delta_1^{K\pi}(s)= {\rm atan} \left({\rm Im} \Gamma_{11}(s) / {\rm Re \Gamma_{11}(s)}\right)
\quad \quad {\rm for} \,\,(M_K+M_\pi)^2 < s < \, (1.6~{\rm GeV})^2~.
\label{eq:phasedisplow}
\end{equation}
At higher energy the phase is unknown. However the knowledge of the asymptotic behaviour of the form 
factor \cite{Lepage} for large $s$ allows to model it in a very rough way. 
Indeed  the phase $\delta_1^{K \pi}$ should go to $\pi$ (modulo $2 \pi$) at 
large $s$. Furthermore sum rules have to be fulfilled, see next section.
Thus  the following simple
model for the phase will be used:  
\begin{equation}
\delta_1^{K\pi}(s)= n_v \pi  \quad \quad  {\rm for} \,\, s > \, (1.6~{\rm GeV})^2~,
\label{eq:phasedisphigh}
\end{equation}
where the quantity $n_v$  should be such that  the sum rules  discussed in 
Sec.~\ref{Sec:sumrules} are satified to a good accuracy.
$\alpha_1$ and $\alpha_2$ in Eq.~(\ref{eq:ffvdisp}) are related to the slope
$\lambda^{'}_+$ and the curvature $\lambda^{''}_+$ of the form factor as 
obtained from Eq.~(\ref{Gam_eq})  
\begin{equation} 
\alpha_1 = \lambda^{'}_+,    \quad \quad \quad \alpha_2 =\lambda^{''}_+- 
\lambda^{'2}_+~.
\end{equation}
%Note that this method allows to easily match to the low energy region where knowledge
%from the vector form factors comes from $K_{\ell3}$  experiments. 
The formula Eq.~(\ref{eq:ffvdisp}) relies on the assumption that the
form factor has no zeros. A technique to find regions on the real axis and
in the complex plane where zeros are excluded has been developped and
applied in particular to the vector and scalar form factors, see \cite{Abbas:2010ns}
for more discussions.  Also the role of zeros in form factors has been 
discussed in \cite{Oller:2007xd}.

\subsection{Scalar form factor}
In the scalar case the inelasticities set in later than in
the vector case \cite{lass,Estabrooks:1977xe}. Therefore, the validity of a single-channel treatment 
is accordingly extended, and it is thus possible, in our region of interest, to
write an expression similar to the one we have just
written assuming  that the form factor has no zeros but with a  simple  single-channel expression for the  phase. A recent discussion on the presence or
absence of zeros in this form factor can be found in \cite{
Doring:2013wka}.
However, in that case, it is more  appropriate to use other subtraction points
than the ones at zero momentum transfer.  
One subtraction is done at zero and the two others at the Callan-Treiman 
point $\Delta_{K \pi}= M_K^2 -M_\pi^2$.
Indeed the Callan-Treiman low-energy theorem \cite{CT} fixes
the value of the scalar form factor at that particular point in the
$SU(2) \times SU(2)$ chiral limit 
\begin{equation}
f_0(\Delta_{K \pi})=\frac{F_K^+}{F_\pi^+} + \Delta_{CT},
\end{equation}
where $F_{K,\pi}$ are the kaon and pion decay constants, respectively,
and $ \Delta_{CT} \sim {\cal O}(m_{u,d}/ 4 \pi F_\pi)$ is a small
correction which has been computed in the framework of $\chi {\rm{PT}}$.
One thus has 
\begin{equation}
\label{Ftil3sub2res}
 f_0^{K\pi}(s) \,=\, f_+(0) \exp\Biggr\{  \frac{s}{\Delta_{K \pi}} \biggr (\ln C +
(s - \Delta_{K \pi}) \alpha  + \tilde G(s) \biggr) \Biggr\}
\end{equation}
with 
\begin{equation}
\tilde  G(s)=\frac{ \Delta_{K \pi} s (s-\Delta_{K \pi} )} {\pi}\!
\int\limits^{\infty}_{s_{K\pi}} \!\!ds'\, \frac{\delta_0(s')}
{s'^2(s'-\Delta_{K \pi} )(s'-s-i\epsilon) } \,,
\end{equation}
and
\begin{equation} 
\alpha=\frac{\ln C}{\Delta_{K \pi}} -\frac{\lambda_0}{M_\pi^2},
\end{equation}
where $\lambda_0$ is the slope of the scalar form factor and  $\delta_0(s)$ its phase.  According to Watson theorem,  $\delta_0$ should coincide 
in the elastic region ($s < \Lambda^2$) with $\delta_0^{K\pi}(s)$, the S-wave I=1/2 $\pi K$ scattering one. Following Refs.~\cite{Bernard:2006gy,Bernard:2009zm} 
one has:
\bea
%\label{eq:phasehighs}
 \delta_0&=&\delta_0^{K\pi}(s) \quad  {\rm{for}} \quad (M_K + M_\pi)^2 <s< \Lambda^2
\nonumber\\
\label{eq:phasehighs}
&=& n_s\pi   \quad \, \, \, \,  {\rm{for}}  \, \, \,\quad s > 
\Lambda^2 \, ,
%\label{eq:phasehighs}
\ena
where $\delta_0^{K\pi}(s)$ is  taken from the work \cite{Buettiker:2003pp}
where a matching of the solution of Roy-Steiner equations with $K \pi \to 
K \pi$, $\pi \pi \to K \bar K$ and  $\pi \pi \to \pi \pi$ scattering data available 
at higher energies was performed. We refer the reader to that work where the
resulting phase $\delta_0^{K\pi}(s)$  is discussed. In 
Eq.~(\ref{eq:phasehighs}) 
$n_s$ can again be estimated such that the sum rules
discussed below are satisfied to a good accuracy. One could also think of
using independent means to constrain this quantity as for example the QCD sum 
rules for the strangeness-changing scalar correlation function which allows to
 relate the strange quark mass to the strange scalar form factor \cite{jop1}.
 While this is
beyond the scope of this paper, investigation
along this line is in progress {\footnote {I would like to thank the referee
for pointing out this fact to me}}. The value of $\Lambda$ will be discussed in Section \ref{sec:combfit}.  

Since there are sum rules which link the two parameters $\ln C$ and $\alpha$
to the high-energy phase we will rather make the fits in what follows with 
a twice-subtracted relation as in \cite{Bernard:2006gy,Bernard:2009zm} 
and study the dependence of our results on the high-energy phase. Thus our
final expression for the scalar form factor will be
\begin{equation}
%\label{Ftil3sub2res}
 f_0^{K\pi}(s) \,=\, f_+(0) \exp\Biggr\{  \frac{s}{\Delta_{K \pi}} \biggr (\ln C  + G(s) \biggr) \Biggr\} \, ,
\label{eq:disptwo}
\end{equation}
with 
\begin{equation}
  G(s)=\frac{ \Delta_{K \pi} (s-\Delta_{K \pi} )} {\pi}\!
\int\limits^{\infty}_{s_{K\pi}} \!\!ds'\, \frac{\delta_0(s')}
{s'(s'-\Delta_{K \pi} )(s'-s-i\epsilon) } \,,
\end{equation}
and the phase defined in Eq.(\ref{eq:phasehighs}).

%Another interesting subtraction point is the soft kaon ananlog of the Callan-Treiman
%point. Instead of taking two subtractions  xxxThis lead to the formula:
%\begin{equation}
%\label{Ftil3sub1res}
% F_0^{K\pi}(s) \,=\, \exp\Biggr\{  \frac{t}{\Delta_{K  \pi}} \biggr(\ln C +
%\frac{(t - \Delta_{K \pi})}{2   \Delta_{K \pi}} (\ln C  -\ln \tilde C) + G(s)\biggr) \Biggr\}
%\end{equation}
%with
%\begin{equation} 
%G(s)=\frac{ \Delta_{K \pi}(t^2-\Delta_{K \pi}^2)}{\pi}\!
%\int\limits^{\infty}_{s_{K\pi}} \!\!ds'\, \frac{\delta_0^{K\pi}(s')}
%{s' (s'-\Delta{K\pi}} \,,
%\end{equation}

\subsection{Sum rules}
\label{Sec:sumrules}
As $t \to -\infty$, one expects $f(t) ={\cal O}(1/t)$
\cite{Lepage}. This asymptotic behaviour dictates the following 
sum rules for the slope and the curvature of the form factors. 
For the vector form factor one has
\bea
\lambda_+' &=&  \frac{ m_\pi^2}{\pi}\!
\int\limits^{\infty}_{s_{K\pi}} \!\!ds'\, \frac{\delta_1^{K\pi}(s')}
{s'^2 } \, \, ,
\label{eq:sumrulev1}
\\
\lambda^{''}_+- 
\lambda^{'2}_+&=&\frac{2 m_\pi^4}{\pi}\!
\int\limits^{\infty}_{s_{K\pi}} \!\!ds'\, \frac{\delta_1^{K\pi}(s')}
{s'^3 } \, \, .
\label{eq:sumrulev2}
\ena
Similar relations hold for $\ln C$ and $\lambda_0$:
\bea
\ln C&=& \frac{ \Delta_{K \pi}} {\pi}\!
\int\limits^{\infty}_{s_{K\pi}} \!\!ds'\, \frac{\delta_0^{K\pi}(s')}
{s'(s'-\Delta_{K \pi} )}
\label{eq:sumrules1}
\\
\frac{\ln C}{\Delta_{K \pi}} -\frac{\lambda_0}{M_\pi^2}&=&
 \frac{ \Delta_{K \pi}} {\pi}\!
\int\limits^{\infty}_{s_{K\pi}} \!\!ds'\, \frac{\delta_0^{K\pi}(s')}
{s'^2(s'-\Delta_{K \pi} )}
\label{eq:sumrules2}
\ena
The sum rule for $\ln C$ has been studied in Ref.~\cite{Bernard:2009zm}.

%In figure \ref{xxx} we show these quantities as a function of the cut-off
%value. 

\section{Results}
\label{Sec:Results}
\subsection{Parameters and their order of magnitudes} \label{Sec:ordermagn}

One has 18 parameters to be fitted in the scattering case: 9 of them corresponds to 
 the  mass and the two coupling constants $g_V^n$ and $\sigma_V^n$  of the three
resonances $K^*(892)$, $K^*(1410)$ and  $K^*(1680)$. The remaining parameters
are the five $a_i$'s
and the four $l_{ab}$, $L^r_{ab}$ with $(a \, b)= K \, \pi$
and $K^*  \, \pi$, respectively.
In the case of  $\tau$ decay 7 parameters more have to be fitted, $\ln C$, 
$I_k^\tau$, $f_+(0) |V_{us}|$, 
$H_2(0)/f_+(0)$ and $F_n/f_+(0)$, the couplings of the three resonances to the
vector source. Note that these couplings as well as $H_2(0)$ are divided by the
value of the vector form factor at zero momentum transfer since  this
quantity cannot be determined as it always  enters combined with $V_{us}$.  
Furthermore since  
our vector  
form factor is only valid up to $\sim 1.65$~GeV we will not integrate  $I_k^\tau(s)$  up 
to $m_\tau$ but rather use  $I_k^\tau$ as a parameter of the fit. We will also
allow the parameter $n_s$ in Eq.~(\ref{eq:phasehighs}) to be free in order to study
the dependence of our results on the high energy region.

Typical order of magnitudes for these parameters are:

$\bullet$ within $\chi$PT one has at leading order $a_0=1/(32 \pi F(3)^2)$ where
$F(3)$ is the pion decay constant in the SU(3) chiral limit.  
Usually $F(3)$ is traded with $F_\pi$, the difference being of higher orders and thus $a_0 =1.16$ GeV$^{-2}$. However there are some indications 
from several studies that 
%that $f_0$ is  smaller than  $f_\pi$ by a non negligible amount
%leading to expect that  $a_0$ could be even twice as large.
%Indeed several studies indicate 
possibly significative differences of patterns exist between the $N_f=2$ and 
$N_f=3$ chiral limits \cite{Moussallam:1999aq}-\cite{Bernard:2010ex}.
%\cite{Moussallam:1999aq,,DescotesGenon:2003cg,DescotesGenon:2007ta,DescotesGenon:2000di,Bernard:2010ex}.
Such differences can be interpreted as a paramagnetic suppression of chiral order parameters when the number
of massless flavors in the theory increases, in relation with the role of $\bar
s s$ vacuum pairs in chiral dynamics. Consequently $F(3)$ could be  smaller 
than  $F_\pi$ in a non negligible way, a ratio $F_\pi/F(3) \sim 1.3$ being not 
excluded. 
We will thus leave 
$a_0$ free in the fit, expecting its value in the range $1.16-2.25$. 

$\bullet$ As stated before  $l_{K \pi}$ and $L_{K \pi}$ contain  contributions from  the polynomial part of the $K \eta$, $K \pi$ loops as well as the tadpoles together with some 
subleading LEC contributions. Typical order of magnitudes for the  ${\cal O}(1/N_c)$ LECs in $\chi$PT at a  scale $m_\rho$ is $10^{-4}$.

$\bullet$ The combination $a_0 + a_1 (m_K +m_\pi)^2$ is related to the $K \pi$ scattering length $a_1^{1/2}$. Indeed expanding the $T$ matrix at small momentum one has
\be
\frac{2}{\sqrt s} T = q_{K \pi}^2( a_1^{1/2} +b_1^{1/2} q_{K \pi}^2 +c_1^{1/2} q_{K \pi}^4 +{\cal O}(q^6))
\en
Some values obtained in the literature for $a_1^{1/2}$ are summarized in 
Table~\ref{tab:pred}. 
%ChPT  and RchPT at ${\cal O}(q^4)$ lead to $ a_1^{1/2}  10 m_\pi^3=0.16(3)$ 
%and  $ a_1^{1/2}  10 m_\pi^3=0.18(3)$ respectively while a Roy-Steiner dispersive analysis of $\pi K$ scattering gave  $ a_1^{1/2}  10 m_\pi^3=0.19(1)$, $ b_1^{1/2}  10^2 m_\pi^5=0.18(2)$ and  $ c_1^{1/2}  10^3 m_\pi^7=0.71(11)$. 
%The results of the $\tau$ decay analysis \cite{xxx}  are  $ a_1^{1/2}  10 m_\pi^3 \sim  0.17$, $ b_1^{1/2}  10^2 m_\pi^5 \sim 0.26$ and  $ c_1^{1/2}  10^3 m_\pi^7 \sim 0.9$.

$\bullet$ Based on the extended NJL model:  $g_V^1 \sim 0.08$ and $\sigma_V^1 \sim
0.25$. 

$\bullet$ As we have seen previously $F^*_K g_V M_K^2/f_\pi^2 \sim 1$ in 
R$\chi$PT thus one expects
$F_K^* \sim 10^{-2} /g_V$, and from our previous estimate $F_K^* \sim 0.1$.

$\bullet$ The value of $H_2(0)$ has been discussed in \cite{mouss}. In the 
chiral limit flavor symmetry is exact and $H_2(0)$ can be related to the
radiative width of the charged $\rho$ meson. Using the experimental value of 
this width  leads to $H_2(0)=(1.54 \pm 0.08)$~GeV$^{-1}$ where the sign was fixed using a vector dominance picture which gives $H_2(0)$ in terms of
the ABJ anomaly.  Refining this estimate taking into account the breaking
of flavor symmetry the author of Ref.~\cite{mouss} 
obtains  $H_2(0) \sim (1.41 \pm 0.09 -65.4 a)\,$GeV$^{-1}$ with $a$
such that $|a| < 10^{-2}$. 

$\bullet$ As we discussed in the introduction  we will add the constraints 
from $K_{\ell 3}$ decays on the values of $\ln C$, $\lambda^\prime_+$  and $f_+(0) |V_{us}|$ which are given in Table~\ref{tab:pred}. One more constraint comes from the branching ratio, Eqs.~(\ref{eq:bkpi}, \ref{eq:bkpiexp}).

\subsection{$K \pi$ amplitude}
In order to determine the parameters which enter the T matrix we will
do a fit to the LASS data \cite{lass}. However  the data available from this
collaboration are given  before unfolding the mass resolution \cite{Dun2012}. Taking this effect into account
affects  significantly the central value of the width of the $K^*(892)$.
 Indeed before unfolding the value is 56 MeV to be compared with the $50$ MeV
result quoted in the literature. However the effect
on its mass value should be very small as well as on the data points above
1 GeV.

Thus in the following we will use the LASS data from 1 GeV to 1.65 GeV.
However
we need the I=1/2 amplitude
since our  aim is to
combine the knowledge from $K \pi$ scattering with the one from $\tau$ decay.
The LASS data being a combination of the I=1/2 and I=3/2 amplitudes
 we will correct them using the following parametrization for $\delta^{I=3/2}$
which is valid above 1 GeV \cite{mousspri}
\be
\delta^{I=3/2}=\arctan(\alpha q_{K \pi}^{3}/(1+\beta q_{K \pi}^6))  \, ,
%\,\, \,\alpha=-0.101292 \pm  0.02121\, , \,\,\, \beta=0.331824 \pm 1.668
\en
where $\alpha=-0.101292 \pm  0.02121$ GeV$^{-3/2}$ and $\beta=0.331824 \pm 1.668$ GeV$^{-3}$ are obtained from a fit to the Estabrooks data \cite{Estabrooks:1977xe}.

\renewcommand {\arraystretch }{1.3}
\begin{table}[b!]
%\label{table:param}
\begin{center}
%%\begin{minipage}
\begin{tabular}{|c|c|c|c|}
\hline
\multicolumn{3}{|c|}{ 
combined fit $\tau$ + $\pi K$}  & \\
\cline{2-3}
& $\lambda^{\prime \prime}_+$ not constrained &  $\lambda^{\prime \prime}_+$  constrained & Exp.\\ 
\hline
 $n_s$&$ 0.788\pm 0.258$&$ 0.785 \pm 0.194$ &  \\
\hline 
%&&&\\
$ \ln C$ \ & $0.2062 \pm 0.0089$ & $0.2064 \pm 0.0081$ &$0.2004 (91)$        \\
&&& $0.2038 \pm0.0241$ ,  $0.1915 \pm0.0116$ \\
&&&$0.1354 \pm0.0133$  , $0.2084 \pm0.0134$ \\
$ f_+(0)| V_{us}|$&$0.2163 \pm 0.0014$ &$0.2163 \pm 0.0012$ &$0.2163(5)$ \\
$ I_k^\tau $ & $0.485 \pm 0.011$ & $0.485 \pm 0.003$ & \\
\hline
%&&&\\
$ \overline{F_1}$ \ &$0.1668 \pm 0.0138$  & $0.1559 \pm 0.000$  &       \\
$\overline{F_2} $\ &$-0.0048 \pm 0.0234$  & $0.0224 \pm 0.001$  & \\
$\overline{F_3}$&$-0.0464 \pm 0.0057$ &$-0.0351 \pm 0.0001$ & \\
$\overline{H_2}(0)$ &$1.46\pm 0.61$ &$1.52\pm 0.02$ &  $H_2(0)$\, : $1.41 \pm 0.09 -65.4 a$ \\ 
%&&&\\
\hline
\end{tabular}
\end{center}
\vspace{0.25cm}
%\hspace{0.25cm}
\begin{tabular}{|c|c|c|}
\hline
%&&\\
$M_1$  &$0.898 \pm 0.013$ &$0.909 \pm 0.000$ \\
$g_V(1)$ & $0.048 \pm 0.006$  & $0.048 \pm 0.000$  \\
$\sigma_V(1)$&$0.334 \pm 0.067$&$0.238 \pm 0.001$ \\
$M_2$ &$1.292 \pm 0.059$ & $1.314 \pm 0.003$  \\
$g_V(2)$ &$-0.015\pm 0.006$ &$-0.0137\pm 0.000$ \\
$\sigma_V(2)$&$0.807 \pm 0.166$ &$0.764 \pm 0.007$  \\
$M_3$ &$1.544 \pm 0.031$  &$1.544 \pm 0.004$ \\
$g_V(3)$& $0.007\pm 0.002$ & $0.007\pm 0.000$ \\
$\sigma_V(3)$&$0.433 \pm 0.053$ &$0.406 \pm 0.010$ \\
%&&\\
\hline
\end{tabular}
%\hspace{-0.18cm}
\begin{tabular}{|c|c|c|}
\hline
%&&\\
$a_0$ [GeV$^{-2}$] &$2.190 \pm 0.132$&$2.270^*$ \\
$a_1$ [GeV$^{-4}$] &$0.067 \pm 0.180$ &$0.054 \pm 0.009 $ \\
$ a_2$ [GeV$^{-2}$]&$-0.187 \pm 2.670$&$-5.807 \pm 0.001$ \\
$a_3$ [GeV$^{-4}$] &$5.122 \pm 1.112$&$6.348 \pm 0.026$ \\
$a_4$ [GeV$^{-3}$] &$-0.308 \pm 0.609$ &$-0.776 \pm 0.009 $\\
$l_{K \pi}\times 10^{-3}$  &$0^*$&$0.093 \pm 0.001$ \\ 
$ L^r_{K \pi}\times 10^{-3}$ &$0.566 \pm 0.141$&$0.560 \pm0.009 $ \\ 
$l_{K* \pi}\times 10^{-3}$ &$0.037\pm 0.264$  &$0.511 \pm 0.001$\\
$ L^r_{K* \pi}\times 10^{-3}$&$0.624 \pm 0.574$&$0.005 \pm 0.003$ \\
%&&\\
\hline
%&&\\
$\chi^2/{\rm d.o.f}$ &$121.26/128$ &$120.70/129$ \\
%&&\\
\hline
\end{tabular}
%\caption{Parameters of the vector form factors}
%\end{center}
%\end{table}
%\end{minipage}
\vspace{0.3cm}
%\captionof{table}{
\caption{Parameters of two combined fits to the  $\tau\to K\pi \nu_\tau$, and $\pi K$ 
scattering data  using some constraints from $K_{\ell 3}$ decays, see text.
In the second
column of the three tables 
the curvature of $f_+(s)$ is unconstrained 
while in
the third it is forced to be within a given range, Eq.~(\ref{eq:chi2const}). 
A bar on a quantity denotes that the quantity is divided
by $f_+(0)$ while a star on a number indicates that the parameter
has been fixed in the fit. The last column of the upper table
gives the corresponding experimental results, the first number for $\ln C$ and the one for $f_+(0) |V_{us}|$ being from $K_{\ell 3}$ 
data taken from the compilation \cite{Flavia2} and the numbers on the second and third line 
for $\ln C$ are in order from KLOE \cite{KLOE}, KTeV \cite{KTeV}, NA48 \cite{NA48} and ISTRA+, see \cite{Flavia2}. The experimental number for $H_2(0)$ is  from \cite{mouss} with 
$|a|<10^{-2}$ where $a$ is a measure of flavor symmetry breaking. 
The  $L^r$'s are evaluated at the scale $\mu=0.897$ GeV and  $\Lambda=1.52$ GeV 
has been used here. The masses of the resonances are in GeV.
}
\label{table:param} 
\end{table}

The data in the elastic region i.e. below 1 GeV can be very well described by the Breit-Wigner form: 
\be
A(s) =\frac{m_{K^*}^2 \Gamma_{K^*}(m_{K^*}^2)}{s-m_{K^*}^2+ i m_{K^*} \Gamma_{K^*}(s)} F_1(s)
\en
with 
\be 
\Gamma_{K^*}(s)= \Gamma_{K^*}(m_{K^*}^2) r  \frac{m_{K^*}}{\sqrt s} F_1^2(s) \, , \quad \quad F_1(s)= r B( q_{K \pi})/B( q_{K \pi}(s=m_{K^*}^2))
\en
where $r= q_{K \pi}/q_{K \pi}(s=m_{K^*}^2)$ and 
$B=1/\sqrt{(1+ r^2_{BW} q_{K \pi}^2)}$ is the Blatt-Weisskopf damping factor. 
%and $r_{BW}=3.40$ GeV$^{-1}$. 
This form reproduces e.g. the LASS data below 1~GeV \cite{lass}. 
There exists, however, more recent results  from the FOCUS collaboration 
\cite{Link:2005ge} based on the 
$D^+ \to K^- \pi^+ \mu^+ \nu$ decay:  
\bea
M_{K^*(892)}&=&895.41 \pm 0.32^{+0.35}_{-0.43} 
\, {\rm {MeV}}  ,  \quad \Gamma_{K^*(892)}=47.79 \pm0.86 ^{+1.32}_{-1.06} \, {\rm{MeV}} \, ,
\nonumber \\  
 &&  \quad \quad \quad r_{BW}=3.96 \pm 0.54 ^{+1.31}_{-0.90} \,\rm{GeV}^{-1} \,.
\ena
\label{eq:lassval}
Data for the phase which could in principle be extracted from this decay \cite{Anan} is not 
available from this collaboration.
We thus  generated our own data below $\sim 1 $ GeV via Monte Carlo. A fit of these data  
leads to  $M_{K^*}=895.41 \pm 0.68$
MeV, $\Gamma_{K^*}=47.80 \pm 1.77$ MeV and $r_{BW}=3.91 \pm  1.86$ GeV$^{-1}$
which is a good representation of the  FOCUS results.  

%We performed a fit of  the 
%so corrected LASS data adding the 
%scattering length as a further contraint on the parameters of the fit. We asked that $a_1^{1/2}$ lies between the  lowest CHPT value and $0.xx$.  The result of the fit is shown in figure~\ref{fig:pik} and 
%the  values for the 18  parameters described in xxx are given in table~\cite{xxx} for the 18  parameters described in xxx. The fit is very good with a $\chi^2=xxx$.

We can now turn to a combined  description  of $\tau$ decays  and $\pi K$ scattering.

%On the LHS the valueof the
%scalar form factor at the Callan-Treiman point $\ln C$, the slope of vector form 
%factor $\lambda_+$, the product of the vector form factor at zero momentum transfer $ \, \, f_+(0))$ and $| V_{us}|$ and $I_k$ the phase space integral. The second column corresponds to the result
%of the fit and the third one to the $K_{\ell3}$ results for $\ln C$ and
%$\lambda_+$ and the global analysis \cite{xxx} for $f_+(0))| V_{us}|$.  On the RHS  the resonance couplings to the
%weak current and the value of the $K^* \pi$ form factor at zero momentum transfer, these four quantities being normalized to the vector form factor at zero momentum transfer }
%\end{minipage}

%\vspace{-2.2cm}
%\hspace{8.2cm}
%\begin{minipage}

\subsection{Combined fit}
\label{sec:combfit}
The $\tau  \to K\pi \nu_\tau$ decay has been measured by Belle and BaBar. Here 
we will fit  the Belle spectrum \cite{Belle} {\footnote {We would like
to acknowledge D. Epifanov for providing us with the Belle spectrum.}}. One
has in the i-th bin
\begin{equation}
N_{\rm events}={\cal N}_T b_w \frac{1}{\Gamma_\tau B_{K \pi}} \frac{d \Gamma_{K\pi}}{d \sqrt t}
\end{equation}
with ${\cal N}_T$ the total number of observed signal events, $b_w$ the chosen bin-width (in GeV/bin) and ${d \Gamma_{K\pi}}/{d \sqrt t}$ the decay spectrum defined in
Eq.(\ref{eq:tauspeckpi}). $\Gamma_\tau$ represents the total
decay width of the $\tau$ lepton and $B_{K \pi}$ is the total branching
fraction, Eq.~(\ref{eq:bkpi}). Clearly,  $f_+(0) |V_{us}|$ appears in this
formula both in the numerator and in the denominator and thus drops from the 
ratio, its knowledge being unnecessary for fitting the spectrum.

\begin{figure}[t!]
\hspace{-0.5cm}
\begin{minipage}{.46\linewidth}
\includegraphics[width=6.5cm,angle=-90]{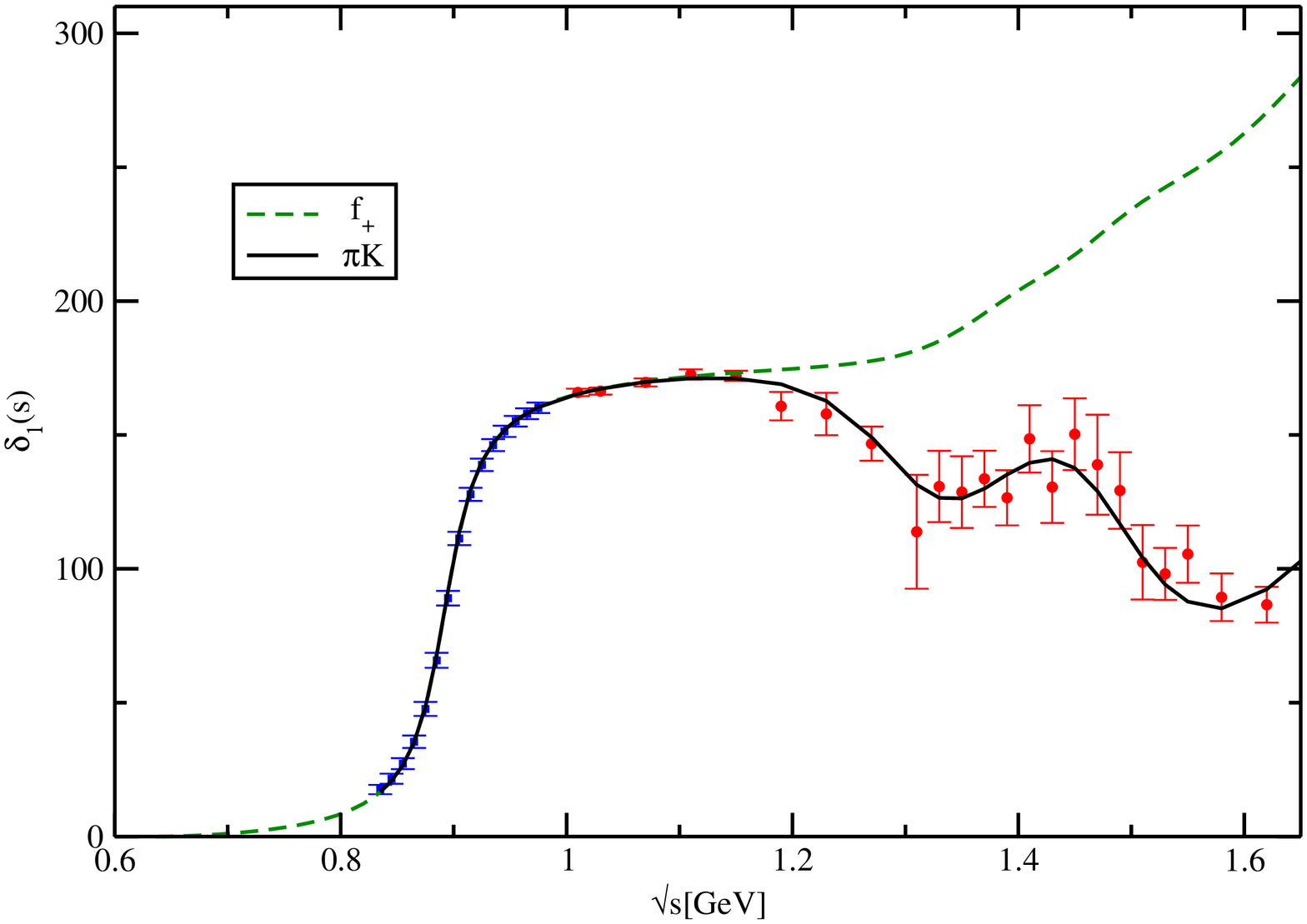}
\end{minipage}
\hspace{0.5cm}
\begin{minipage}{.46\linewidth}
\includegraphics[width=6.5cm,angle=-90]{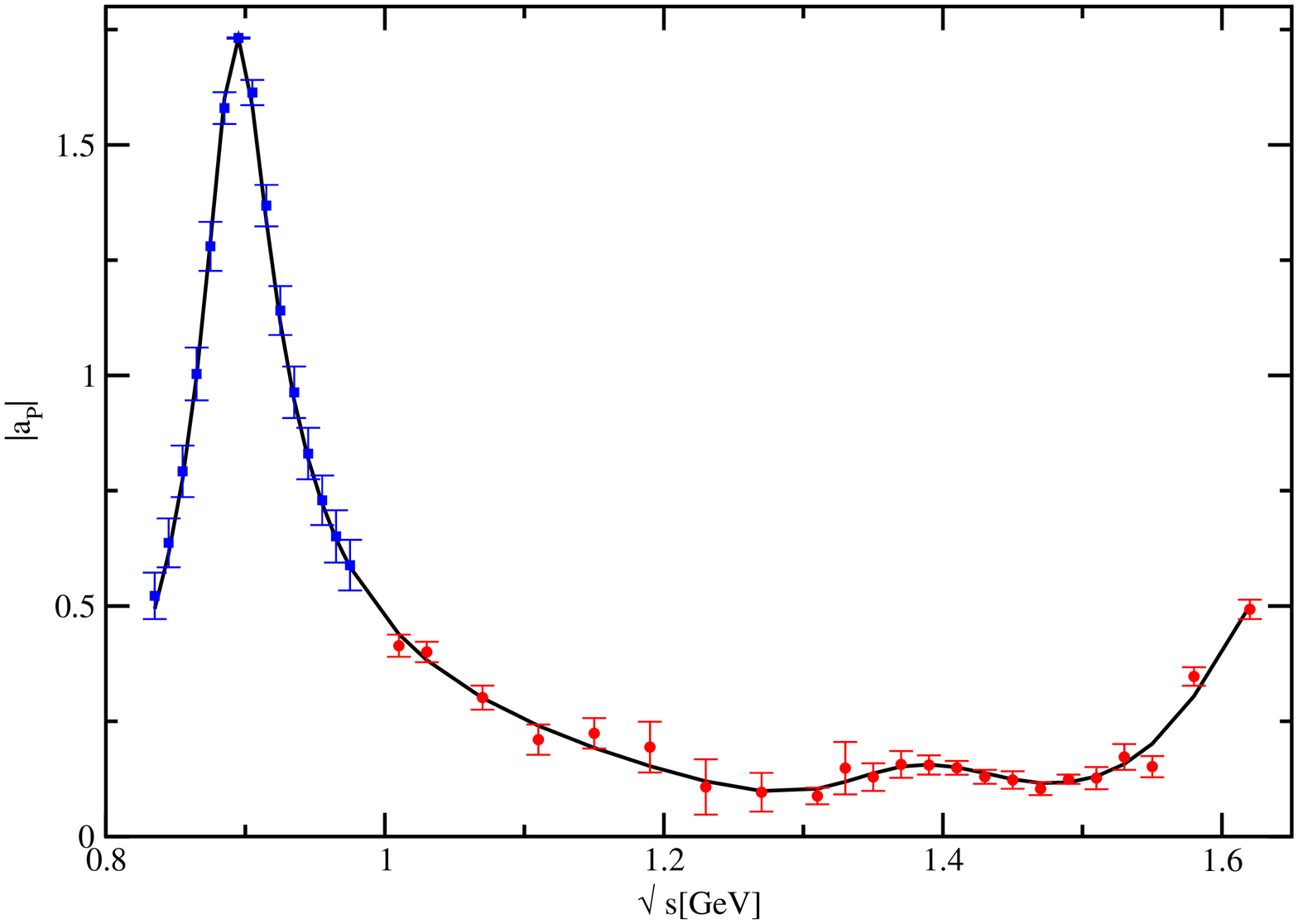}
\end{minipage}
\caption{Left panel: Phase of the vector form factor (green dashed line) 
compared to the $P$ wave $I=1/2$ $\pi K$
phase (black solid line). Right panel: Modulus of the $P$ wave $I=1/2$ $\pi K$ amplitude. The blue
squares are the data generated via Monte Carlo using the FOCUS results and the
red circles are the corrected LASS data, see text. }
\label{fig:pik}
\end{figure}

\begin{figure}[t!]
\hspace{0.5cm}
\begin{center}
\includegraphics[width=10cm,angle=-90]{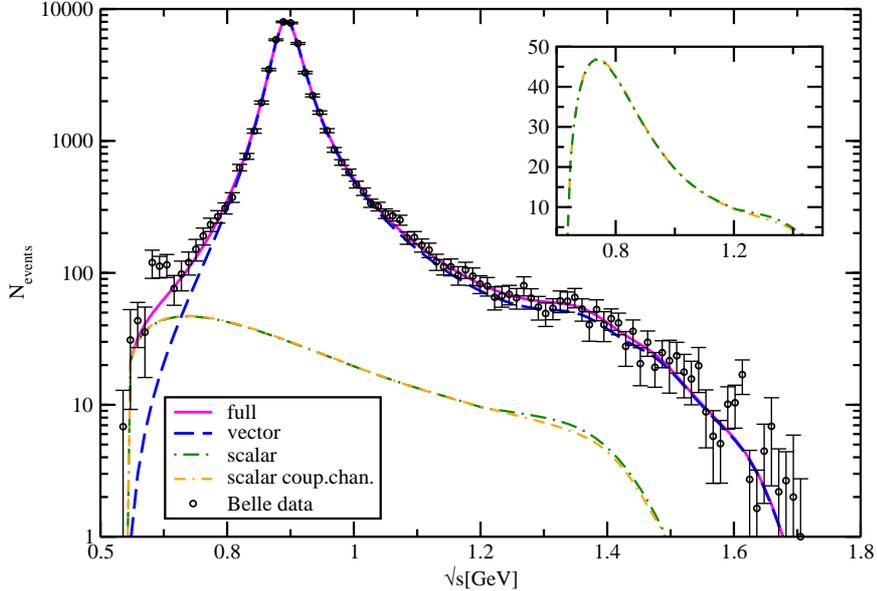}
\end{center}
\caption{Spectrum of $\tau\to K\pi \nu_\tau$. The black circles are  the Belle
data \cite{Belle}. The dot-dashed green and the dot-dashed-dashed orange line are the scalar form factor 
contribution from the dispersive analysis, Eq.~(\ref{eq:disptwo}) and the 
coupled channel model \cite{jop,ElBennich:2009da} respectively. 
The dashed blue line represents the vector form factor contribution and the 
solid magenta line gives the full result. The inset shows the scalar form 
factor contribution on a linear scale. }\label{fig:tau}
\end{figure}

\begin{figure}[t!]
\hspace{0.5cm}
\begin{center}
\includegraphics[width=8.8cm,angle=-90]{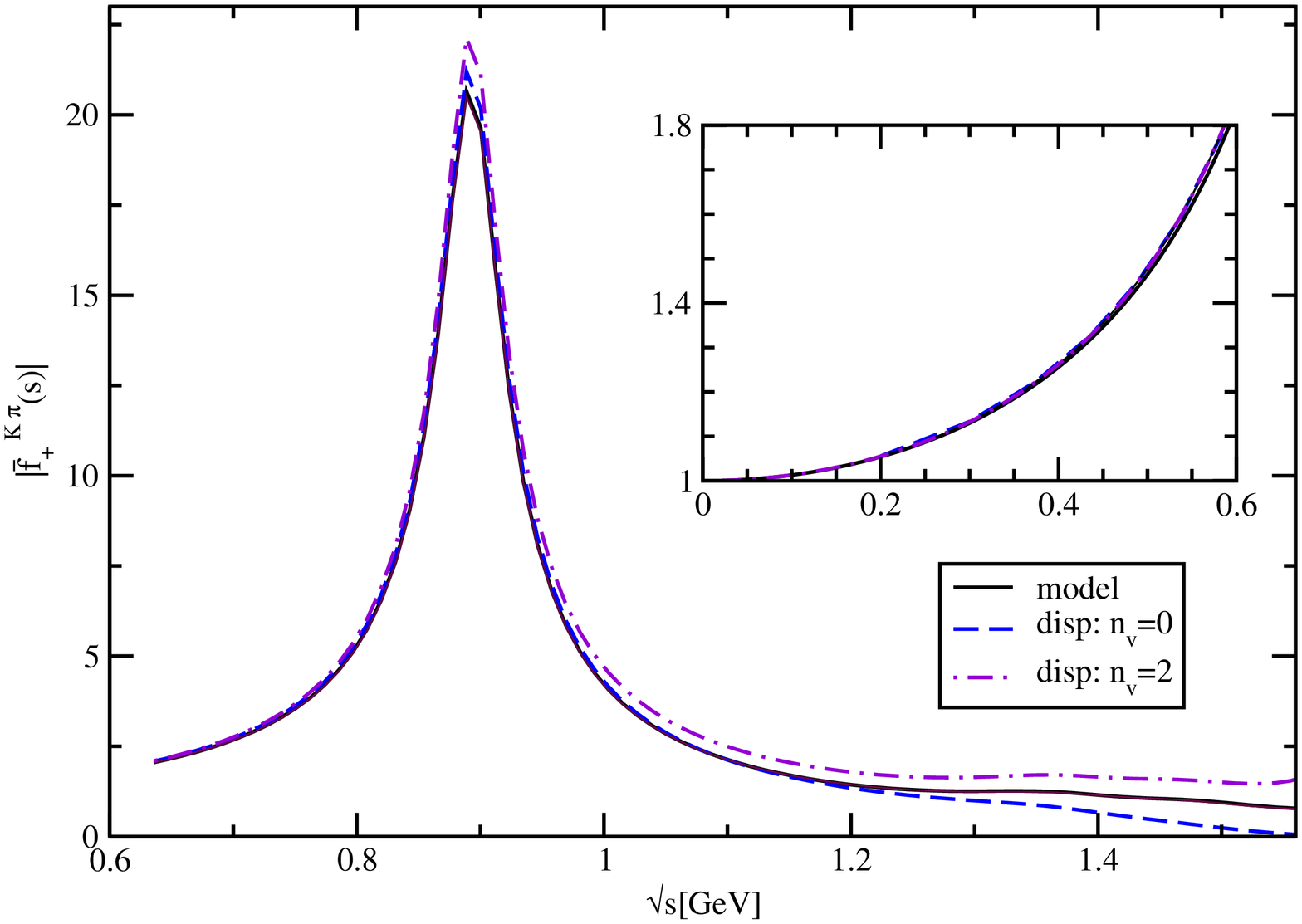}
\end{center}
\caption{Modulus of the normalized vector form factor from the combined fit using the model parametrization of the vector form factor compared to the three subtracted dispersive analysis for two
different values of the
parameter $n_v$, Eq.~(\ref{eq:phasedisphigh})  which parametrizes  our ignorance of the phase at high energy.
The solid black line corresponds to the result of the combined fit. The inset
shows the result for low $\sqrt s$, the three curves in that region are almost
undistiguishable.}
\label{fig:vecff}
\end{figure}

\subsubsection{Fit with constraint on $f_+(0) |V_{us}|$}
Table~\ref{table:param}
gives the value of the
$\chi^2/{\rm {d.o.f}}$ and  of the parameters obtained from a combined 
fit to $\tau\to K\pi \nu_\tau$ and $\pi K$ scattering data 
with some contraints from  $K_{\ell3}$ decays and from the newest value of $B_{K \pi}$
\cite{Ryu:2014vpc}. We will first discuss the case without constraint
on the curvature
of the vector form factor. 
%In the following,  we will 
%concentrate on the 
%results with the constraint from  the newest value of $B_{K \pi}$ 
%\cite{Ryu:2014vpc} and only comment when the differences between
%the two cases become somewhat sizeable. 
Fig.~\ref{fig:pik} compares respectively
the $I=1/2$ $\pi K$ phase and modulus of the amplitude in the $P$ wave  with the corrected LASS data above the elastic region and the 
data generated from the FOCUS results below. The number of events 
$N_{\rm events}$ from the Belle data and the model is depicted in Fig.~\ref{fig:tau} as a 
function of $\sqrt s$. It is clear from these figures that the combined fit is 
excellent. This is confirmed by the very good $\chi^2$/d.o.f defined as:

\bea
\chi^2&\equiv&\chi^2_{{\rm noct}\lambda^{\prime \prime}}= \chi^2_{{\rm {noctfV}}} + \left(\frac{|f_+(0) V_{us}| - |f_+(0) V_{us}|^{\rm exp}}
{\sigma_{|f_+(0) V_{us}|^{\rm exp}}}\right)^2 \, ,
\nonumber \\
\chi^2_{{\rm {noctfV}}}&=& \sum_i \left(\frac{\delta_i - \delta_i^{\rm exp}}
{\sigma_{\delta_i^{\rm exp}}} \right)^2 + 
\sum_i \left(\frac{a_i - a_i^{\rm exp}}
{\sigma_{a_i^{\rm exp}}} \right)^2 + \left(\frac{a_1^{1/2}-a_1^{\rm exp}}
{\sigma_{a_1^{\rm exp}}} \right)^2 +
\sum_i \left( \frac{N_i- N_i^{\rm exp}}
{\sigma_{N_i^{\rm exp}}} \right)^2 \nonumber \\
&& 
+ \left( \begin{array}{cc} 
\ln C & -\ln C^{K_{\ell 3}} \\
\lambda'_+& -\lambda'_+{^{K_{\ell 3}}}  \end{array} \right)^T V^{-1}
 \left( \begin{array}{cc} 
\ln C & -\ln C^{K_{\ell 3}} \\
\lambda'_+& -\lambda'_+{^{K_{\ell 3}}}  \end{array} \right) 
+ \left(\frac {B_{K \pi} - B_{K \pi}^{\rm exp}}{\sigma_{B^{\rm exp}}}\right)^2
\nonumber \\
&&
%+
%\left(\frac{|f_+(0) V_{us}| - |f_+(0) V_{us}|^{\rm exp}}
%{\sigma_{|f_+(0) V_{us}|^{\rm exp}}}\right)^2  
+\left(\frac{n_s -0.75}{0.25}\right)^2 + \left(\frac{h_2-H_2(0)}{0.75}\right)^2
\label{eq:chi2} 
\ena
where V is the covariance matrix and $\rho(\ln C, \lambda'_+)=-0.33$ 
\cite{Flavia2}. Our fit
is performed with  84 points
from the Belle data  in the energy region from threshold to 1.6 GeV 
%\footnote 
%{Since 
%the data are rather bad as one gets closer to the $\tau$ mass we restricted the fit
%to $\sqrt s =1.6$ GeV} 
and
39 experimental points  $\delta_i^{exp}$ for the phase of $\pi K$ scattering up to 1.66 GeV. 
Since in the elastic region phase and amplitude are related via a sinus we 
only 
fit the amplitude above this region and thus one has only 24 data points 
$a_i^{exp}$ in  
the second sum.  We have constrained 
$f_+(0) |V_{us}|$  as $0.2160 \pm 0.0014$ which corresponds to the error 
band given by the $K_{\ell 3}$ data without averaging them but 
rather taking the smallest/largest value obtained in the various
experiments, see discussion 
below.  Furthermore $ B_{K \pi}^{\rm exp}= 0.416 $ and $\sigma_{B^{\rm exp}}=0.008$, see Eq.~(\ref{eq:bkpiexp}). The former smaller result for $B_{K \pi}$, Eq.~(\ref{eq:bkpi}) leads to similar
results with essentially somewhat smaller values for $I_k^\tau$ and  the 
curvature of the vector form factor. 
 We will come back on the constraint on $n_s$ below. The last constraint
 takes into account the fact that $h_2$ is a leading order result and thus should 
dominate if one expects the series to converge rapidly. The same holds of course for $h_1$,
but we did not enforce it in the fit.  
One gets $h_1=1.04$ and $h_2 =1.35$.
The various terms contribute to the $\chi^2$ as follows:
\be
 \chi^2= 40.2 + 80.6 + 0.42 + 0.07 + 0.03  +0.02   + 0.02   
\en 
where the first number corresponds to the sum of the three first terms (17.8 + 17.5   
+4.8) in Eq.~(\ref{eq:chi2}) i.e. it measures the quality of the fit of 
$ \pi K$ scattering. 
Note that in the Belle data, Fig.~\ref{fig:tau} there is a bump close to threshold given
by three points, bins 6,7 and 8 which cannot be accomodated within
our parametrization (as well as others) and which does not seem to be
present either in the BaBar data \cite{BaBar} or in the more recent 
Belle data \cite{Ryu:2013lca}. This region contributes for 27
to the $\chi^2$, so that without these points the latter would
be even better. In  Fig.~\ref{fig:tau} is also shown the contribution to the
spectrum from the scalar form factor and the vector one. The former
clearly dominates in the threshold region, the vector one being responsible
for the peak at the $K^*(892)$ resonance. A measurement of the forward-backward asymmetry
would be very useful to disentangle the two contributions \cite{Truong}, helping
to get a better precision on the parameters of the two form factors.  In  Fig.~\ref{fig:pik} the phase
of $\pi K$ scattering in the $P$ wave is identical to the one of the vector form factor in the 
elastic region as demanded by the Watson theorem and starts to deviate 
when the inelasticities set in.    

The parameter $\Lambda$, Eq.~(\ref{eq:phasehighs}) is set to $1.52$ GeV in this fit. We have
also performed a fit with $\Lambda=1.67$ GeV leading to similar
results. Thus we refrain to show them here.  Indeed the value where the inelasticities set in in the $S$ wave is not
very well known. A reasonable range of values is $1.43 \, \rm{GeV} < \Lambda < 1.67$
GeV where the lower value is determined by the $K^*$ resonance and the upper
one  is the energy where the phase of
the amplitude is experimentally found to be different from the phase
of the $S$ matrix. Some discussion related to this can be found for example
in \cite{Bernard:2009zm,mouss}.

\begin{figure}[t!]
%\label{fig:kstarpirs}
\hspace{-0.5cm}
\begin{minipage}{.46\linewidth}
\includegraphics[width=6.5cm,angle=-90]{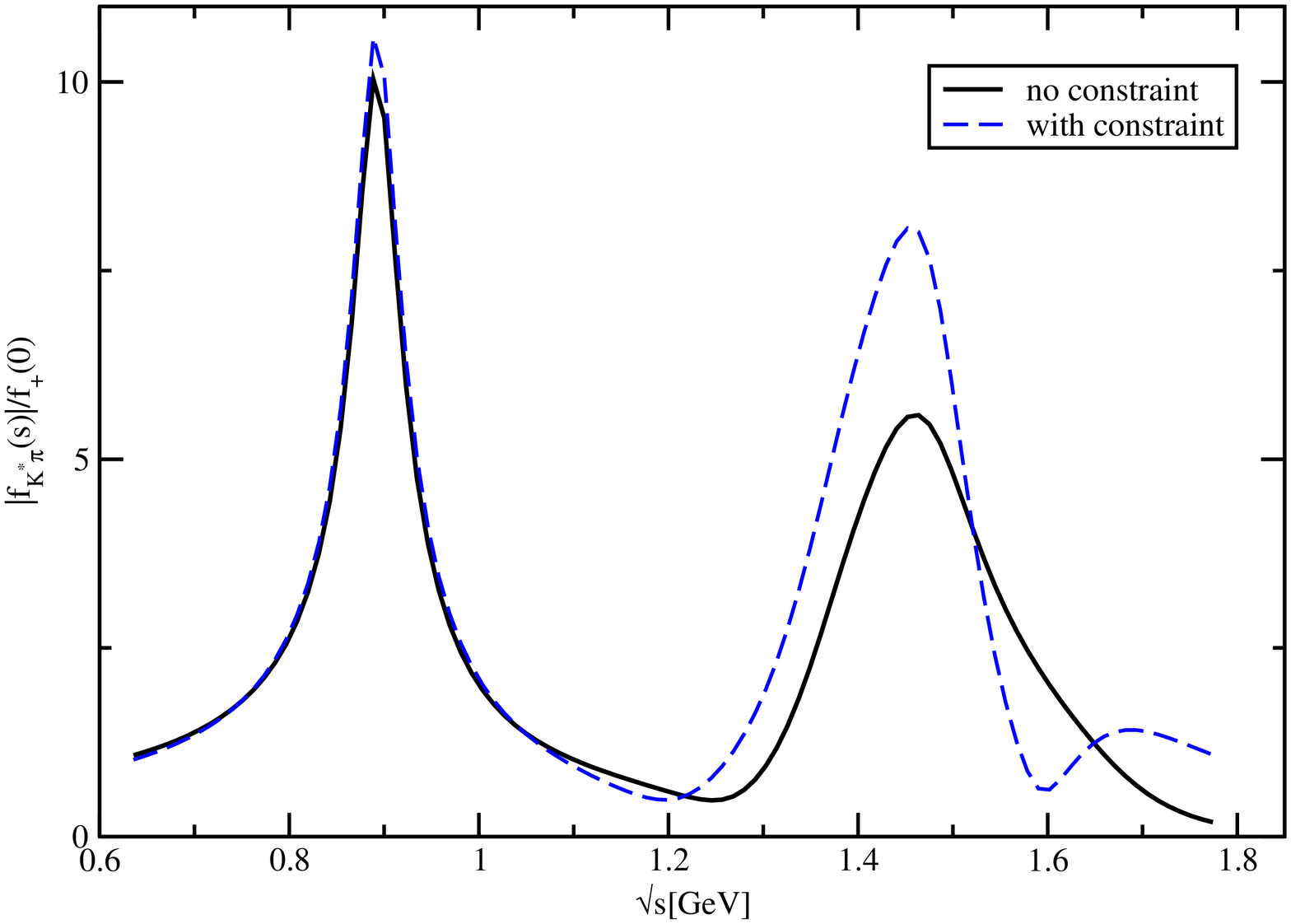}
\end{minipage}
\hspace{0.5cm}
\begin{minipage}{.46\linewidth}
\includegraphics[width=6.5cm,angle=-90]{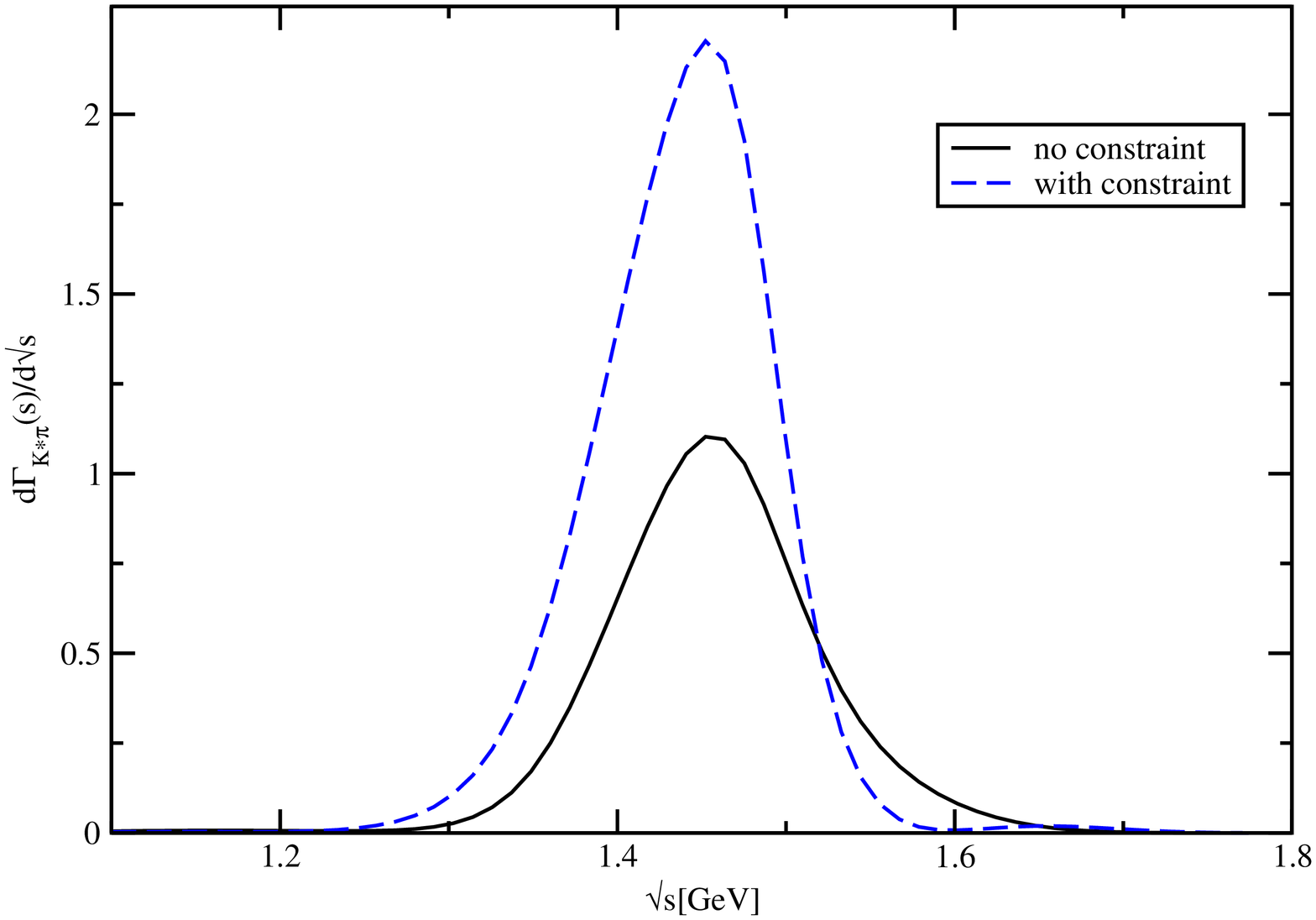}
\end{minipage}
\caption{Modulus of the $K^* \pi$ form factor divided by $f_+(0)$ (left panel) and energy distribution of the decay width (right panel) as obtained in 
the combined fit. The black solid and blue dashed lines  correspond to 
the fit without and with constraint on the curvature of the vector form 
factor respectively. }
\label{fig:kstarpirs}
\end{figure}

The values of our parameters in Table~\ref{table:param}
are compatible with the estimated order of magnitudes
discussed in section \ref{Sec:ordermagn}. 
$a_0$ is at the upper end of the expected range leading to a rather 
small value of the decay constant in the chiral limit in favor of 
a paramagnetic suppression of the pion decay constant in the SU(3) chiral 
limit compared to the SU(2) one. 
 $g_V(1)$/$\sigma_V(1)$ are respectively 
somewhat smaller/larger 
than the ENJL predictions of Ref.~\cite{Prades}.  $H_2(0)/f_+(0)$ 
compares well with  its experimental value ($f_+(0)$ is typically between 0.95 and 1), 
leading to a very small flavor breaking value $a$.  Note
also that integrating the spectrum obtained from the fit
gives a value of $I_k^\tau$ consistent with the  value
determined by the fitting procedure. 
We will discuss the results for $f_+(0) |V_{us}|$ and $\ln C$ below.
Few remarks concerning the masses of the resonances are in order. First
these are model dependent quantities. Second the data from LASS and FOCUS
concern the neutral $K^*$  while the published Belle analysis 
correspond to the charged $K^*$. Here we did not take into account isospin
breaking however the PDG gives a difference of about 4 MeV between the 
two masses. Following \cite{bjp,bjp1} we have thus calculated the complex pole 
positions $s_R = m_R^2 - i \Gamma_R m_R$ \cite{Escribano:2002iv} in the second
Riemann
sheet of the vector resonances which are much less model dependent. 
It also allows to determine the width of these resonances.  
One gets
\bea
&M_{K^*(892)}=(891.29 \pm 7.7)
\, {\rm {MeV}} \quad  \quad, \quad   &\Gamma_{K^*(892)}=(46.26 \pm 5.01) \, {\rm{MeV}}
\nonumber
\\
&M_{K^*(1410)}=(1370.65 \pm 35.93)
\, {\rm {MeV}} \, , \quad   &\Gamma_{K^*(1410)}=(164.93 \pm 34.56) \, {\rm{MeV}}
\label{mwreso}
\ena
%As in the work \cite{xxx} whose results for the $K^*(892)$ are in very good agreement
%with ours,
As expected from the quality of the fit,
the results for the $K^*(892)$ are in agreement with LASS, Eq.~(\ref{eq:lassval}) within the error bars   while the  central value of the mass  is close to the PDG recommended 
value 
$M_{K^*(892)}=891.66 \pm 0.26$~MeV for the charged $K^*$.
The  width is somewhat too small though within the error bars, the PGD quotes  
for the charged $K^*$, $\Gamma_{K^*}=50.8 \pm 0.9 \, {\rm{MeV}}$.  However as noted in \cite{bjp1}  the PDG values
are chiefly obtained from the parameters of Breit Wigner type expressions and thus need not to be exactly the same as determined from the pole position. This remark also holds for the $K^*(1410)$ where the
PDG gives $M_{K^* (1410)}= 1414 \pm 15  {\rm{MeV}}$ and $\Gamma_{K^*}=232 \pm 21 \, {\rm{MeV}}$.

In Fig.~\ref{fig:vecff} is shown the result of the fit for the modulus of the
normalized vector form factor. It
increases from one at zero momentum transfer up to the $K^*(890)$ resonance 
region where it shows a strong peak. 
The values of its slope and curvature are given in Table~\ref{tab:pred} and compared with  results obtained from a quadratic fit to  $K_{\ell3}$ data and 
various
theoretical results from earlier works on $\tau \to K \pi \nu_\tau$ decay.
The slope is in good agreement
and the curvature even though  compatible with most of the experimental
results  which
are rather spread with large error bars, has  a central value  
a bit 
small compared to the  theoretical results. Also our results for the slope
and curvature lie within the allowed domains obtained in \cite{Abbas:2010ns,Caprini:2010ye} within
the method of unitarity bounds.
% If we perform a quadratic fit of our form factor
%in the region of $K_{ell3}$ decay we get $25.61$ and $1.24$
The vector form factor is compared in Fig.~\ref{fig:vecff} to the result of a dispersive
analysis, Eqs.~(\ref{eq:ffvdisp}-\ref{eq:phasedisphigh}) for two values of the parameter $n_v$,
corresponding to  a very conservative
estimate of  our ignorance of the
phase at high energy, using our result for 
the slope and the phase up to 1.6 GeV, while the curvature is determined from 
the sum rule, Eq.~(\ref{eq:sumrulev2}). The generated band  
is very small up to $\sim 0.85$ GeV and broadens as the energy increases 
further. However the uncertainty 
from the high energy phase is not too large up to 1.6~GeV. Our form factor 
is compatible with the dispersive analysis for $n_v \sim 0$.
%within the band as it should if the form factor has no zero since it has the good properties of analyticity and crossing. 
Let us consider the sum rules.   
The RHS of the first one, 
Eq.(\ref{eq:sumrulev1}) is
$(17.022 + 7.609 \, n_v) \times 10^{-3}$  where the first number corresponds to
the integral from threshold to  $\Lambda=1.6$ GeV while the second is the
remaining contribution up to infinity
taking the value $\pi$ for the phase. As expected the latter contribution is 
sizeable leading in principle to a rather large uncertainty from the
high energy region.  
%Taking on the LHS the value
%of the slope obtained in the fit leads to $n_v=1.12$.
For a not too large violation of the sum
rule $n_v$ should lie typically between 0.74 and 1.4. Clearly the second sum rule,
Eq.~(\ref{eq:sumrulev2})
has a much smaller uncertainty from the high energy region, one gets from the RHS 
 $(5.556 + 0.579 \, n_v) \times 10^{-4}$ leading to a value of the curvature of the
form factor using $n_v$ in the range just given and taking into account the
error on the slope, $1.23  \times  10^{-4} <\lambda^{\prime \prime} _+ < 1.30 \times 10^{-4} $. We will briefly come back on the issue of the size of
$\lambda^{\prime \prime} _+$ at the end of the section.

The modulus of the $K^* \pi$ vector form factor is illustrated in the left panel of
Fig.~\ref{fig:kstarpirs}. It has two peaks of the same order of magnitude, one at the  $K^*(890)$ resonance  which is of course
much less pronounced than the analog peak in $f_+(s)$ and the other one close
to the second  resonance $K^*(1410)$. This  leads to the energy distribution of the decay width
${d\Gamma_{K^*\pi}(s) / d\sqrt{s}}$, Eq.~(\ref{eq:gamkstar}) shown in the right panel of the same 
figure. It 
is consistent with the theoretical work \cite{mouss}. Integrating this 
distribution gives 
the integrated rate $R(\tau
\to K^*(1410) 
\nu_\tau \to K \pi \pi \nu_\tau) $. The result is shown in Table~\ref{tab:pred}.  The central 
value is smaller than the Aleph result, however,    
within the error bars which are rather large both
for theory and experiment. 
Upcoming experiments on  $\tau
\to K^*(1410) 
\nu_\tau \to K \pi \pi \nu_\tau$  will help constraining the parameters
of the model further.

\begin{table}[t!]
\begin{center}
%\begin{minipage}
\begin{tabular}{|c|c|c|c|}
\hline
%& & \\[-0.1cm]
$a_1^{1/2}$ &&& \\[-0.4cm]
%\times  10 \, m_\pi^3
& $0.249 \pm 0.011$&$0.247 \pm 0.001$ &$0.16(3) \, \, 0.18 \, , \,0.18(3)  \, , \,
0.19 (1)  \, ,\, 0.17 $ \\[-0.5cm]
$\times  10 \, m_\pi^3$&&& \\[0.25cm]
\hline
%& & \\[-0.1cm]
%$\lambda'_+ $&$25.58 \pm 0.38 $ &$25.58 \pm 0.40 $
$\lambda'_+ $& & &$20.64(1.75) \, , \,25.6(1.8) \, , \,
24.86 (1.88)  \, , \, 24.80(1.56)  $ \\[-0.4cm]
&$25.56 \pm 0.40 $ &$25.58 \pm 0.09 $& \\[-0.35cm] 
 $\times  10^3$& & & $26.05_{-0.58}^{+0.21} \, , \, 25.20(33)  \, , \,  24.66(77) \, , \,  25.49 (36)  $ \\[0.25cm]
\hline
%& & \\[-0.1cm]
%$\lambda^{\prime \prime} _+ \times 10^3$&$ 0.97 \pm 0.06$&$ 1.10\pm 0.12$   
$\lambda^{\prime \prime} _+$& & &$3.20(69) \, , \,1.5(8)  \, , \,
1.11 (74)  \, , \, 1.94(88) $\\[-0.4cm]
&$ 1.11 \pm 0.08$&$ 1.22\pm 0.02$   &
\\[-0.35cm]
$ \times  10^3$& & &$ 1.29_{-0.04}^{+0.01} \, , \, 1.29(3)  \, , \, 1.20(2) \, , \,  1.22(2)$
\\[0.25cm] 
\hline
%& & \\[-0.1cm]
$ B_{K \pi} [\%]$ & $0.414 \pm 0.008 $&$0.414 \pm 0.005$ &$0.404 \pm 0.02 \pm 0.013$ 
%\cite{Belle} 
$\, , \, 0.416 \pm 0.01 \pm 0.008$ 
%\cite{Ryu:2014vpc} 
\\[0.25cm]
\hline
%& & \\[-0.1cm]
$R \times 10^3$ &$0.70 \pm 0.43 $&$1.23 \pm 0.05$ &$1.4^{+1.3}_{-0.9}$\\[0.25cm]
\hline
\end{tabular}
\end{center}
\vspace{0.25cm}
\caption{Prediction for the  $K \pi$ scattering
length $a_1^{1/2}$, the slope and
curvature of the vector form factor, the branching ratio and the integrated rate $R(\tau
\to  K^*(1410) 
\nu_\tau \to K \pi \pi \nu_\tau) $. The second and third column give
respectively the results of the fit without and with the constraint on
the curvature of the vector form factor.   
%\cite{Belle}   and \cite{Ryu:2014vpc}, see the last column for the values 
%quoted in 
%these papers in that order. 
The last column  summarizes also various theoretical
predictions for $a_1^{1/2}$,  $\lambda'_+$ and $\lambda^{\prime \prime}_+$ as
well as experimental results for the two latter quantities  and the integrated 
rate.  From left to right the numbers for $a_1^{1/2}$ correspond to  $\chi$PT at ${\cal O}(p^4)$ \cite{Bernard:1990kw} and 
at ${\cal O}(p^6)$ \cite{Bijnens:2004bu}, R$\chi$PT   at ${\cal O}(p^4)$ \cite{Bernard:1990kw1}, a Roy-Steiner dispersive
analysis of $\pi K$ scattering \cite{Buettiker:2003pp} and a $\tau$ decay 
analysis \cite{bjp1}. The experimental numbers from $K_{\ell 3}$ data (first line) for $\lambda'_+$ and $\lambda^{\prime \prime}_+$ are from left
to right from  KTeV \cite{Alexopoulos:2004sy} , KLOE 
\cite{KLOE,Ambrosino:2006},  NA48 \cite{NA48,Lai:2007} and ISTRA+ \cite{ISTRA}. 
The theoretical numbers (second line) are from earlier works on $\tau\to K\pi \nu_\tau$  without constraints from 
$K_{\ell 3}$ \cite{Jamin:2008qg}-\cite{bjp} and  with constraints \cite{bjp1}.
The experimental results for $B_{K \pi}$ are from \cite{Belle,Ryu:2014vpc} respectively.}
\label{tab:pred}
\end{table}

In Table~\ref{tab:pred} are also given the predicted values for the $K \pi$ scattering
length $a_1^{1/2}$ and  the branching ratio $B_{K \pi}$, Eq.~(\ref{eq:bkpi}). 
$a_1^{1/2}$ turns out to be somewhat too large compared to 
various predictions, the last column giving some results from 
$\chi$PT, R$\chi$PT, Roy equations  and  $\tau$ decay.
Note however that there is a lack of constraints from the experimental data in the
threshold region  and that the same too large value 
was also obtained in a similar combined analysis \cite{mouss} contrary to 
\cite{bjp1} where
only the Belle spectrum was fitted.

Let us discuss the value of $n_s$, Eq.~(\ref{eq:chi2}).
As we have seen in section \ref{Sec:sumrules}  $\ln C$ obeys a sum rule. Using our 
parametrization of the unknown phase, Eq.~(\ref{eq:phasehighs}) one gets 
\be
n_s  = \frac{1}{G_{as}} \left(\ln C - \frac{ \Delta_{K \pi}} {\pi}\!
\int\limits^{\Lambda}_{s_{K\pi}} \!\!ds'\, \frac{\delta_0^{K\pi}(s')}
{s'(s'-\Delta_{K \pi} )} \right)  
=  \frac{1}{0.10446} (0.2062-  0.1336) = 0.696
\en
where $G_{as}$ corresponds to the integral from $\Lambda$ to infinity with the 
phase equal to $\pi$. The sum rule  
is satisfied for $n_s=0.696$. We have allowed for some 
violation of the sum rule since $G_{as}$ is not  known, our fit leading to a $5\%$ 
discrepancy.  As discussed previously for the  vector form factor the second sum rule, 
Eq.~(\ref{eq:sumrules2})
has a much smaller uncertainty from the high energy region, one gets from the RHS of this equation,  
 $0.152 +0.018 \, n_s$. Thus with $n_s$ as given from the fit
the slope of  
the scalar form factor  is
$\lambda_0=0.0144 \pm 0.0007$.
%In the case $\Lambda=1.67\,$GeV one has
%from the first sum rule $n_s = (0.2056 - 0.1512)/0.08534 = 0.637$ while the  fit leads to $n_s=0.767$ 
%which corresponds
%again to a $5\%$ discrepancy for the sum rule.  

\begin{figure}[t!]
%\hspace{0.5cm}
%\begin{center}
%\includegraphics[width=8cm,angle=-90]{fs75.ps}
%\end{center}
\hspace{-0.5cm}
\begin{minipage}{.46\linewidth}
\includegraphics[width=6.5cm,angle=-90]{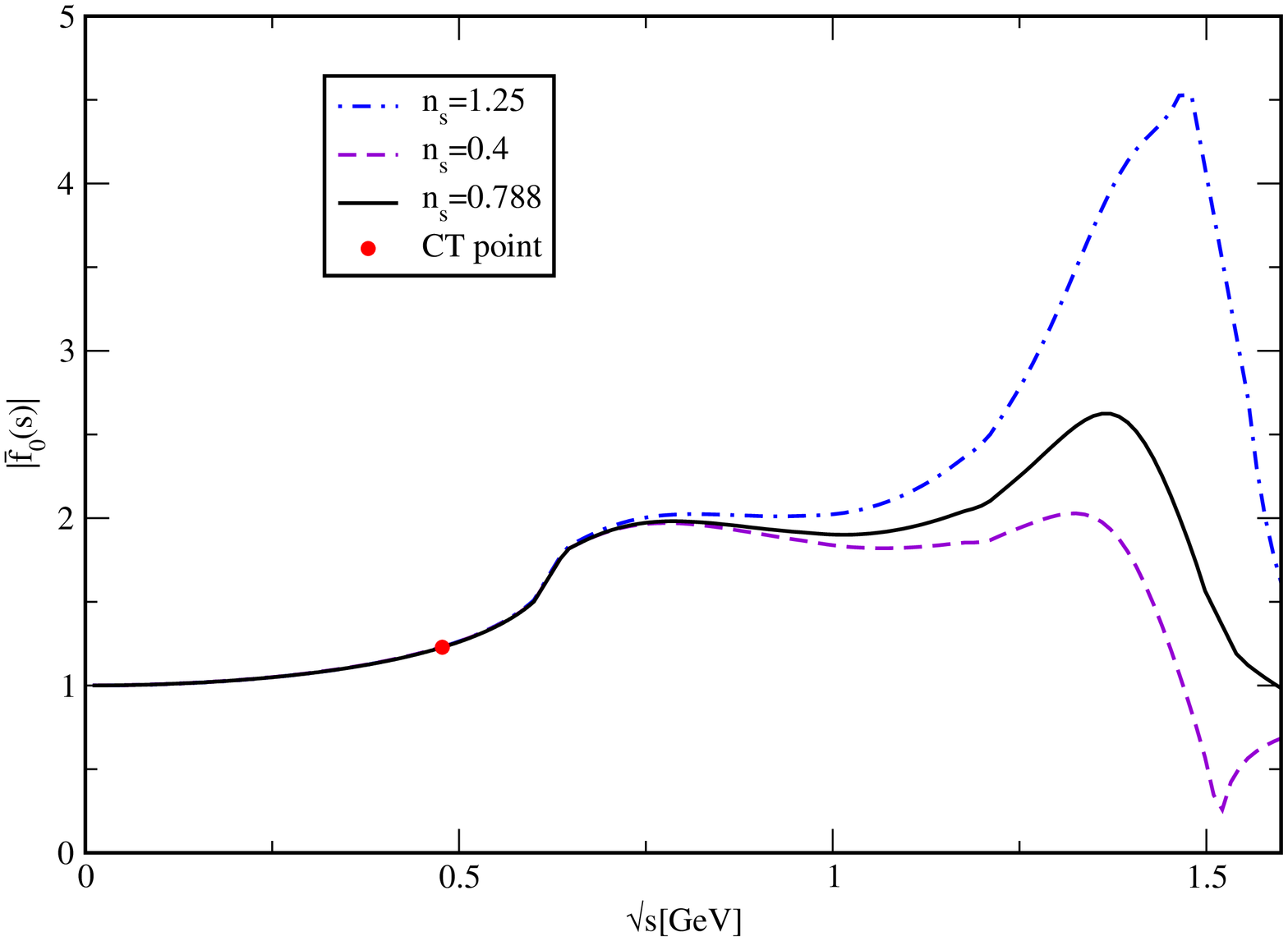}
\end{minipage}
\hspace{0.5cm}
\begin{minipage}{.46\linewidth}
\includegraphics[width=6.5cm,angle=-90]{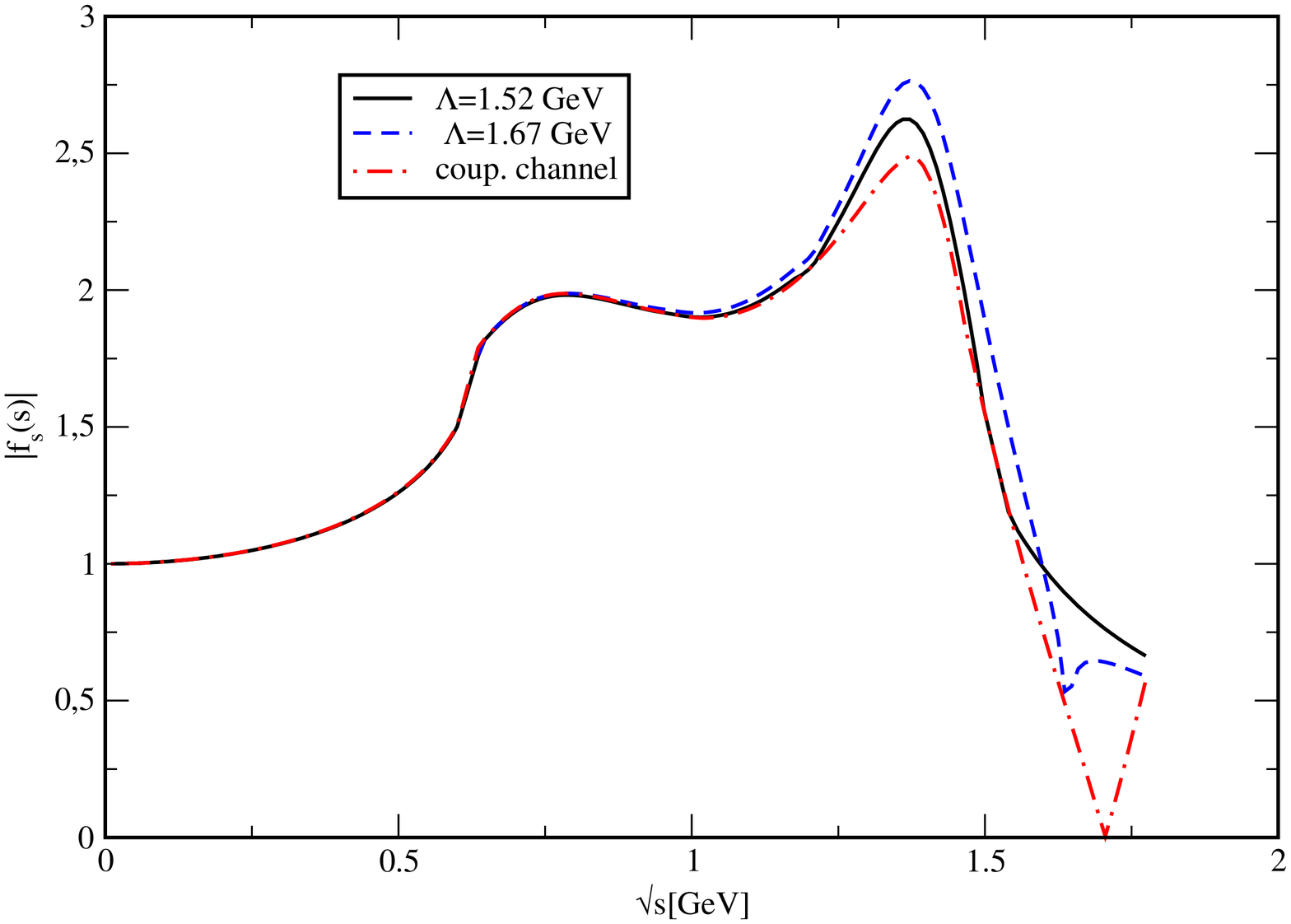}
\end{minipage}
\caption{Left panel: modulus of the normalized scalar form factor for three different values of the
parameter $n_s$, Eq.~(\ref{eq:phasehighs})  which parametrizes  our ignorance of the phase at high energy.
The solid black line corresponds to the result of the combined fit. The Callan-Treiman (CT)
point is shown by the red circle. Right panel: modulus of the normalized scalar form factor
for two different values of the cut off $\Lambda$, Eq.~(\ref{eq:phasehighs}) and from a fit to a coupled channel analysis \cite{jop,ElBennich:2009da}.}
\label{fig:scalff}
\end{figure}

The modulus of the normalized scalar form factor is depicted in Fig.\ref{fig:scalff} for three
different values of the parameter $n_s$ keeping the value at the CT point
fixed.  These values gives a  violation of the sum rules by $15\%$ for
$n_s=0.4$ 
and $30\%$ for  $n_s=1.25$. The uncertainty due to the high energy phase is 
 much larger than in the vector form factor case, fortunately the sum rules 
help reducing it sizeably. The form factor has a first small bump around the $K^*(890)$ resonance and a second one around the $K^*(1410)$
one, the latter being more or less pronounced depending on the value of $n_s$. This behaviour
agrees with older calculations of the $\pi K$ scalar form factor, see \cite{ 
jop} as well as the recent work \cite{Doring:2013wka}.  The $\tau$ data combined with $\pi K$ scattering plus constraints
from the sum rules demand a somewhat stronger second bump compared to the first
one which compares also very well  with 
\cite{jop1}. The behaviour of our form factor above 
$\sim 1.25$ GeV  is  sensitive to
the value of the parameter $\Lambda$ as shown on Fig.~\ref{fig:scalff}.
To compare further our model independent description of the
scalar form factor  we have repeated the combined fit~{\footnote {I would like 
to thank
B. Moussallam for providing me with his fortran code.}}  using
a coupled channel dispersive analysis analogous to \cite{jop,ElBennich:2009da} for describing this form factor. Indeed 
such a model has been extensively used in  various works on $\tau\to K\pi \nu_\tau$. However in the line of what has been done  here we do not fix the value of the
scalar form factor
at the CT point contrary to what is done in these works.  We obtain very similar results for the fit
parameters and thus refrain to present them here, let us just
quote the value of $\ln C=0.2061 \pm 0.0086$, $f_+(0) |V_{us}|$ being the same 
as in Table~\ref{table:param}. The three form factors are  
compared in  Fig.~\ref{fig:scalff} while in  Fig.~\ref{fig:tau} the scalar contributions to $N_{{\rm {events}}}$ obtained in the fit with $\Lambda=1.52\,$GeV and with the coupled channel analysis are shown.   The three
form factors start to differ as one gets closer to the region  
where the inelasticities set in due to a different drop of the phase more or 
less abrupt which is then followed by a growth, see 
\cite{Bernard:2009zm,ElBennich:2009da,Ananthanarayan:2004xy,Oller:2007xd}. 

\begin{figure}[t!]
\begin{center}
\includegraphics[width=6.5cm,angle=-90]{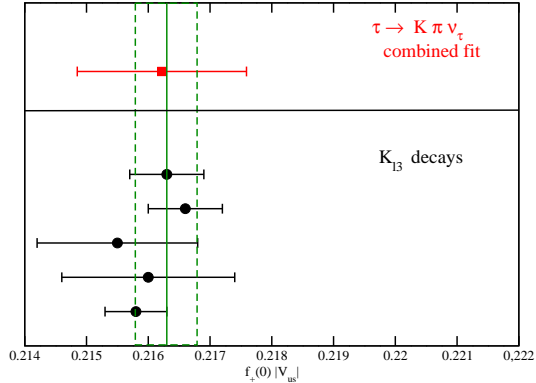}
\end{center}
\caption{Determination of $f_+(0) |V_{us}|$ from semileptonic kaon decays (on the lower portion)
and from the combined fit (upper portion). 
%The two results in the latter corresponds to
%two different constraints on this quantity as explained in the text. 
The band corresponds
to the average of the semileptonic data from \cite{Flavia2}.}
\label{fig:fpvus}
\end{figure}

Let us finally discuss the value of $f_+(0) |V_{us}|$
obtained adding the constraint from $K_{\ell 3}$ decays as explained
below Eq.~(\ref{eq:chi2}). It is compared  in Fig.~\ref{fig:fpvus} to
several values determined from five 
$K_{\ell 3}$ decay modes, see Ref.~\cite{Flavia2} for more details.  
%We  also show the result of a fit
%where we relax the constraint on  $f_+(0) V_{us}$ using 0.21 as central value  and 0.01
%as errors (orange line in the figure).
Assuming the SM couplings  and using $V_{ud}=0.97425 (22)$ from a recent survey
\cite{Hardy:2008} one gets 
\be
f_+(0)=0.959(6) \, ,
\label{eq:fp0fitSM}
\en
which is within the error band of the lattice averaging from FLAG-1 \cite{Colangelo:2010et}, 
$f_+(0)=0.956(8)$. It is also compatible with the results from the updated version 
\cite{FLAG2}. There, according to the FLAG quality criteria,
the results of two collaborations are
given as the new averages:  $f_+(0)=0.9667(23)(33)$ 
from MILC ($N_f=2+1$) \cite{MILC} and $f_+(0)=0.9560(57)(62)$ from ETM09A ($N_f=2$) \cite{ETM09A}.

One can now compare the result for $\ln C$  from the fit with its expression 
from the Callan-Treiman theorem. Experimental  information on the inclusive
$K_{\ell 2}$ and  $\pi_{\ell 2}$ decay rates and precise knowledge of
the radiative corrections lead to \cite{Flavia2}
\be
 F_K^+/F_\pi^+ |V_{us}/ V_{ud}| =0.2758(5) \, .
\en 
Assuming again the SM couplings 
and the value of $f_+(0)$ as given by our fit, Eq.~(\ref{eq:fp0fitSM}) one gets 
\be
\Delta_{CT}=(-1.29 \pm 1.28) \times 10^{-2} \, ,
\en
%whose central value is 
%somewhat larger than expected from an  $SU(2) \times SU(2)$ 
%discrepancy. {\bf XXX why do you say that??? 1 percent is quite good for SU(2)!!
whose central value is somewhat larger than expected from $\chi$PT calculations. However considering the large error bar the value of $\Delta_{CT}$ is compatible  with the NLO
$\chi$PT result in the isospin limit \cite{GL} $ (-0.35 \pm 0.8) \times 10^{-2}$  
(the error is a conservative estimate of the higher order corrections),
NNLO estimates with isospin breaking  given in 
\cite{Kastner:2008,Bijnens:2007},  and chiral extrapolations to 
lattice data \cite{Bernard:2010ex,Bernard:2009ds}. 
%They lead to $\Delta{CT}< 10^{-2}
Similarly for the form factor at the soft kaon analog point one gets
\be
\bar f_0(-\Delta_{K \pi})=0.865 \pm  0.008  \, , \quad \quad  
\tilde \Delta_{CT}=
(-0.972  \pm 0.941 ) \times 10^{-2}  \, ,
\en
where $ \tilde \Delta_{CT}$ is defined by the following SU(3)$\times$SU(3) theorem
\be
f_0(-\Delta_{K \pi})=\frac{F_{\pi^+}}{F_{K^+}} + \tilde \Delta_{CT} \, .
\en
As for
$\Delta_{CT}$ the error bars are large and the result is again compatible 
with $\chi$PT calculations and chiral extrapolation of the lattice data 
\cite{Bernard:2010ex,Bernard:2009ds}. In the former case on gets 
$\tilde \Delta_{CT}=0.03$ at NLO  in the isospin limit \cite{GL},
while at  two loop order two low energy constants enter \cite{bta} leading to the
following estimate $-0.035 <\tilde \Delta_{CT} < 0.11$ \cite{Bernard:2009zm}.
Hence 
at present our results are compatible with the SM.

\begin{table}[t!]
\begin{center}
%\begin{minipage}
\begin{tabular}{|c|c|c||c|c|c|}
\hline
&$\lambda^{\prime \prime}_+$ unconstr.&$\lambda^{\prime \prime}_+$  constr.&&
$\lambda^{\prime \prime}_+$ unconstr.&$\lambda^{\prime \prime}_+$  constr. \\
\hline
$\ln C$ & $ 0.2051  \pm 0.0088$& $0.2043 \pm 0.0081 $ & $f_+(0) |V_{us }|$ &
$0.227 \pm 0.008$& $0.230 \pm 0.002$\\
\hline
$M_K$ & $ 892.70 \pm 1.11 $ & $ 891.62 \pm 0.18 $ &$M_{K^*}$ &$1366.76 \pm 28.44  $ &$1376.24 \pm 2.64 $\\
\hline
$\Gamma_{K}$&$ 46.62\pm 1.11 $ &$46.21 \pm 1.97$ & $\Gamma_{K^*}$&$ 155.72\pm 39.68$& $195.51 \pm 2.37 $\\
\hline
$\lambda^\prime \times 10^3 $ &$25.56 \pm 0.41 $ &$25.59 \pm 0.05 $&
$\lambda^{\prime \prime}\times 10^3 $ &$0.81 \pm 0.30 $ &$1.22 \pm 0.02 $  \\
\hline 
$I_k^\tau$ & $0.444 \pm 0.029$ & $0.431 \pm 0.002$    
 &$R \times 10^3$ &$0.66 \pm 0.41$  &$1.24 \pm 0.04$ \\
\hline
$n_s$ & $0.774 \pm 0.259$&$0.758 \pm 0.204$ & $\chi^2$ &$119.89/128$ &$118.58/128$ \\
\hline
\end{tabular}
\caption{Results of two fits where  $f_+(0) |V_{us}|$ is left free and 
$\lambda^{\prime \prime}$ is either constrained or not. The mass and width of 
the resonances are in MeV. For more details see text and Tables~\ref{table:param},~\ref{tab:pred}.}
\label{table:noctfVus}
\end{center}
\end{table}

\vspace{-0.7truecm}
\subsubsection{Role of the constraint on $f_+(0) |V_{us}|$ and of the 
curvature of $f_+(s)$.}
Before concluding let us discuss the role played by the constraint on 
$f_+(0) |V_{us}|$ from $K_{\ell 3}$ decays which we have 
included in our fit as well as the result on the curvature of 
$f_+(s)$.

In order to see the role played by the constraint on 
$f_+(0) |V_{us}|$  (see Eq.~(\ref{eq:chi2}) and discussion below) 
we have performed a similar fit without this constraint. The results are
shown in Table~\ref{table:noctfVus}.  The $\chi^2 \equiv \chi^2_{{\rm {noctfV}}}$ is similar in the two cases, however the
value of $f_+(0) |V_{us}|$ is larger with a smaller uncertainty. Consequently
the value of $I_k^\tau$ is smaller, the value of $B_{K \pi}$ being similar in the 
two fits
due to the rather strong constraint from the new Belle result.
Interestingly the central value of the curvature of the vector form factor is now much smaller 
leading to a $35 \%$ violation of the sum rule for $n_v=0$. However the error
bar is rather large.  

We have thus performed a new fit constraining the value
of $\lambda^{\prime \prime} _+$ to be within the  range given 
in   \cite{bjp1}, see
Table~\ref{tab:pred}, first leaving the constraint on  $f_+(0) |V_{us}|$.
The $\chi^2$ has now one additional term   
\be
\chi^2= \chi^2_{{\rm noct}\lambda^{\prime \prime}} +
\left( \frac{\lambda^{\prime \prime}_+ - 1.22}{0.02}\right)^2  
\label{eq:chi2const}
\en
where $ \chi^2_{{\rm noct}\lambda^{\prime \prime}}$ is the expression, Eq.~(\ref{eq:chi2}).
%It turns out that there are two possible solutions with similar $\chi^2$,
%one with a negative value for $F_2/f_+(0)$ as in the unconstrained fit and one 
%with a small positive one. They both lead to results comparable to the %ones discussed previously but for the $K^* \pi$ vector form factor in the latter case. 
%We will thus just discuss briefly this case.
In this new fit we fixed the value of  $a_0$ since a 
larger value of the slope prefers a larger value of this parameter. Results  are compared in 
the third columns of Table~\ref{table:param} with the fit without constraints
on the curvature. The values
of $\ln C$, $f_+(0)|V_{us}|$ and $\Delta_{CT}=(-1.26 \pm 1.14) \times 10^{-2}$ are quite stable, the parameters mostly changed being the ones related to
the $K^*\pi$ channel. However due to the rather strong constraint we have
imposed,
the error bars are in most cases much smaller. The mass and width of the resonances are now:
\bea
&\, M_{K^* (892)}=(891.22 \pm 1.70)
\, {\rm {MeV}} \quad \quad \!, \quad   &\Gamma_{K^*(892)}=(46.26 \pm 1.99) \, {\rm{MeV}}
\nonumber\\
&M_{K^* (1410)}=(1379.84 \pm 23.59)
\, {\rm {MeV}} \, ,  \quad  &\Gamma_{K^*(1410)}=(179.35 \pm 36.42) \, {\rm{MeV}}
\ena
to be compared with Eq.~(\ref{mwreso}).
%$H_2(0)$ is somewhat larger leading to 
%a somewhat larger breaking of flavor symmetry $a=0.00x-0.00x$.
Concerning
the predictions, Table \ref{tab:pred},  similar
results are obtained for most of the quantities except the curvature and 
the integrated rate $R$ which is now in better agreement with the central value of Aleph. This is due to a second bump in the modulus of the $K^* \pi$ form factor
which is now more pronounced than in the fit discussed in the previous subsection as seen in Fig.~\ref{fig:kstarpirs}.
As already stated a better
measurement
of the energy distribution of the  decay width  would be very useful to constrain the parameters
of the fit.
%The slope  of the vector
%form factor is also unchanged while the curvature is
%\be  
%\lambda^{\prime} _+=25.57 \times 10^{-3} , \,\,\, \lambda^{\prime \prime} _+ = 1.23\times 10^{-3}
%\en
%The value of the integral of the phase of the vector form factor up to 1.6 GeV,
% Eq.(\ref{eq:sumrulev2}) 
%is $5.693\times 10^{-4}$ and the sum rule
%gives $\lambda^{\prime \prime} _+ =1.21$ for $n_v=0$.
%The modulus of the $K^* \pi$ form factor is compared to the 
%result of our former fit on Fig.~\ref{fig:kstarpirs}. The second bump 
%has now the 
%same height than the first one leading 
%to a larger value for 
%$R=1.189 \times 10^{-3}$ 
%in better agreement with the central value of Aleph. As already stated a better%measurement
%of the energy distribution of the  decay width 
%of $\tau \to K^*(1410) \nu_\tau \to K \pi\pi 
%\nu_\tau $.    

For completeness we have finally repeated the fit without the constraint on $f_+(0) |V_{us}|$. The results
are compared in Table~\ref{table:noctfVus} with the similar fit but without
constraint on the curvature of the vector form factor. 

The main conclusion 
from the studies performed on the role of the constraint on  $f_+(0) |V_{us}|$   
 is that given the experimental uncertainties on the spectrum the combined fits
prefer a  smaller value of $I_k^\tau$ and consequently a larger value of 
$f_+(0) |V_{us}|$, the product of these two quantities being constrained by
the branching ratio. However the values obtained are too large compared
to the $K_{\ell 3}$ ones. Clearly more precise data are needed 
to be able to  determine $f_+(0) |V_{us}|$ from $\tau$ data alone.   
%and would either lead to a rather large value of $f_+(0)
%|V_{us}| \sim 1$ assuming the SM result for $|V_{us}|$ or to a $\sim 5 \%$ deviation of

%The conclusion concerning 
%the comparison of our results with the SM ones is unchanged. 

\section{Conclusion}
The study performed here offers for the first time a direct extraction of 
$f_+(0) |V_{us}|$ from 
$\tau\to K\pi \nu_\tau$ decay.
A  model for the vector form factor  valid in the region below
 $\sqrt s \sim 1.65$ GeV is  build from a N/D
method. Using 
a simple dispersive approach for the scalar form factor (as well
as a coupled channel method for comparison) a
combined analysis of 
$\tau\to K\pi \nu_\tau$ decay 
and  $\pi K$ scattering constrained by $K_{\ell 3}$ and $D_{\ell 4}$ data 
is performed. The coupled channel approach used here for the vector
form factor allows to determine also the decay spectrum of $\tau \to K^*(1410) \nu_\tau
\to K \pi \pi \nu_\tau$ which is at present not very precisely measured.
The result obtained for $f_+(0) |V_{us}|$ is almost 
independent of the model used for the scalar form factor. The value of this 
form factor at the Callan-Treiman point as well as the soft kaon analog  
determined from the fit are compared
to $SU(N_f) \times SU(N_f)$ theorems with $N_f=2$ for the former and $N_f=3$ for the latter. At the level of accuracy of the
data our results are compatible with the Standard Model. However,
the forthcoming experiments will
help reducing the uncertainty 
on $f_+(0) |V_{us}|$ and $\ln C$ allowing  for a stringent test of the
Standard Model. Indeed the errors in the $\tau$ spectrum according to the 
expected sensitivity of a second generation
B factory
% assuming an integrated luminosity of $40$ ab$^{-1}$, see e.g.
%\cite{Abe}
will be considerably reduced allowing for a determination of $f_+(0) |V_{us}|$ 
from $\tau$ data alone.  Futhermore
a measurement of the forward-backward asymmetry would be very useful to 
disentangle  the scalar and vector form factors in the $\tau$ spectrum. 
Finally this analysis should be refined to 
include the long distance electromagnetic and strong isospin breaking
corrections and the effects  from the unphysical cuts
which have been neglected here once a much better precision of the data is reached. 

\acknowledgments

I thank  B. Moussallam for sharing with me his very deep insights
into the subject and S. Descotes-Genon and A. Le Yaouanc 
for enlightening discussions. I would also like to thank D. Boito for
his participation at an early stage of the work and for 
very interesting discussions and E. Passemar for some checks at
an early stage of the work.  I am also grateful to D. Boito, M. D\"oring, U.-G. Mei{\ss}ner, and B. Moussallam
 for careful reading of the manuscript and  U.-G. Mei{\ss}ner for
useful comments. This work is supported in part by the "EU I3HP Study of
Strongly Interaction Matter" under the seventh Framework Program of the
EU.

\end{document}